\documentclass[twocolumn]{aastex631}
\usepackage{makecell}

\begin{document}

\title{A search for self-lensing binaries with TESS and constraints on their occurrence rate}

\author[0000-0001-6970-1014]{Natsuko Yamaguchi}
\affiliation{Department of Astronomy, California Institute of Technology, 1200 E. California Blvd, Pasadena, CA, 91125, USA}

\author[0000-0002-6871-1752]{Kareem El-Badry}
\affiliation{Department of Astronomy, California Institute of Technology, 1200 E. California Blvd, Pasadena, CA, 91125, USA}

\author[0000-0002-3562-9699]{Nicholas M. Sorabella}
\affiliation{Lowell Center for Space Science and Technology, Univ. of Massachusetts Lowell, Lowell, MA 01854, USA}

\begin{abstract}
% 300 words
Five self-lensing binaries (SLBs) have been discovered with Kepler light curves. They contain white dwarfs (WDs) in AU-scale orbits that gravitationally lens solar-type companions. Forming SLBs likely requires common envelope evolution when the WD progenitor is an AGB star and has a weakly bound envelope. No SLBs have yet been discovered with data from the Transiting Exoplanet Survey Satellite (TESS), which observes far more stars than Kepler did. Identifying self-lensing in TESS data is made challenging by the fact that TESS only observes most stars for $\sim$25 days at a time, so only a single lensing event will be observed for typical SLBs. TESS's smaller aperture also makes it sensitive only to SLBs a factor of $\sim$100 brighter than those to which Kepler is sensitive. We demonstrate that TESS has nevertheless likely already observed $\sim$4 times more detectable SLBs than Kepler. We describe a search for non-repeating self-lensing signals in TESS light curves and present preliminary candidates for which spectroscopic follow-up is ongoing. We calculate the sensitivity of our search with injection and recovery tests on TESS and Kepler light curves. Based on the 5 SLBs discovered with Kepler light curves, we estimate that $(1.1 \pm 0.6)\%$ of solar-type stars are orbited by WDs with periods of 100-1000 d. This implies a space density of AU-scale WD + main sequence (MS) binaries a factor of 20-100 larger than that of astrometrically-identified WD + MS binaries with orbits in Gaia DR3. We conclude that the Gaia sample is still quite incomplete, mainly because WD + MS binaries can only be unambiguously identified as such for high mass ratios. 
\end{abstract}


\section{Introduction} \label{sec:intro}

Self-lensing binaries (SLBs) are eclipsing binaries containing a compact object that gravitationally lenses the light of its companion when it transits in front of it. Self-lensing produces a light curve feature that resembles an inverted transit, wherein the binary brightens once per orbit. Like planetary transits, these pulses generally have short durations, small duty cycles, and small amplitudes. Although self-lensing was recognized as a possible avenue to detect compact object binaries more than 50 years ago \citep{Leibovitz1971A&A, Maeder1973A&A}, the first SLB was only discovered a decade ago with the Kepler mission, which provided high-precision, rapid cadence light curves for more than $10^5$ stars \citep{Kruse2014Sci}. Since then, four additional systems have been identified with Kepler \citep{Kawahara2018AJ, Masuda2019ApJL, Yamaguchi2024PASP}. All SLBs discovered so far contain $\sim 0.6\,M_{\odot}$ white dwarfs (WDs) orbiting solar-type stars in AU-scale orbits. Self-lensing also holds promise for the detection of neutron stars and black holes in wide orbits \citep[e.g.][]{Masuda2019ApJ_2, Wiktorowicz2021MNRAS, Chawla2023arXiv}, but such companions are likely rarer than WDs, and none have been discovered via self-lensing to date.

Self-lensing WD binaries both allow for precision measurements of WD masses and radii \citep[e.g.][]{Yahalomi2019ApJ} and the detection of WD companions in relatively wide orbits, which are difficult to detect with most other methods. SLBs are complimentary to the sample of astrometric WD binaries recently discovered via Gaia astrometry \citep{Shahaf2024MNRAS}. AU-scale WD + MS binaries are valuable as probes of binary interactions, having orbital periods of 100 - 1000 days -- intermediate between the traditional predictions of stable and unstable mass transfer products \citep[e.g.][]{Zorotovic2010A&A, Ivanova2013A&ARv, Belloni2024A&A, Yamaguchi2024MNRAS, Yamaguchi2024PASP_2}. SLBs can help to constrain the space density and population demographics for these binaries because they have a relatively simple selection function which  depends primarily on geometry. In contrast, the probability of a wide WD + MS binary being detected via Gaia astrometry is more complicated to model, given the complex cascade of astrometric models and detection thresholds employed in constructing Gaia binary catalogs \citep[e.g.][]{Halbwachs2023A&A}.  

The Kepler mission continuously monitored the same $100$ deg$^2$ patch of the sky for 4 years \citep{Borucki2010Sci, Koch2010ApJL}, during which it observed over $100,000$ stars with 30 minute cadence. In contrast, the Transiting Exoplanet Survey Satellite (TESS), observes a $\sim 2000$ deg$^2$ region for 27 days before moving on to a new region, allowing it to cover almost the entire sky in two years \citep{Ricker2014SPIE}. While this means that TESS will observe many more stars than Kepler, most of these have baselines of just 27 days every two years (though some in regions near the ecliptic poles can have up to $\sim 1$ year of continuous observations). This is disadvantageous to finding long-period systems like the known SLBs, for which Kepler detected multiple pulses. With TESS, we generally expect to detect only one pulse, making it more difficult to filter out false-positives. Moreover, TESS has an aperture $\sim 100$ times smaller than Kepler and thus has much lower sensitivity at fixed brightness. Nonetheless, \citet{Wiktorowicz2021MNRAS} predicted that TESS will discover about $60 - 170$ SLBs.

Several works have since attempted the search \citep{Sorabella2023ApJ, Sorabella2024ApJL}, but to date, no new SLB candidate from TESS has been confirmed. In this work, we describe our search for non-repeating self-lensing pulses in TESS light curves. We also revisit the Kepler-detected SLBs and constrain the occurrence rate of AU-scale WD+MS binaries. In Section \ref{sec:tess_selection}, we describe properties of our initial TESS sample and in Section \ref{sec:main_search}, our method to identify candidate self-lensing signals. In Section \ref{sec:injection_and_recovery}, we describe injection and recovery tests to measure the detection efficiency of our algorithm. Using the results of these tests, in Section \ref{sec:rates}, we calculate occurrence rates of WD companions to solar-type stars in AU-scale orbits. We also compare the occurrence rate to values inferred from the Gaia astrometric sample of WD + MS binaries. Finally, we conclude in Section \ref{sec:conclusion}. Spectroscopic follow-up of our best candidates will be described in future work.

\section{Light curve selection} \label{sec:tess_selection}

The TESS mission collects images of its entire field of view with a cadence of 30 minutes, 10 minutes, or 200 seconds, depending on sector. These full-frame images (FFIs) are calibrated and processed through several different pipelines to extract light curves \citep{Huang2020RNAAS}. In our search, we use PDCSAP light curves provided by the TESS Science Processing Operations Center (TESS-SPOC; \citealt{Jenkins2016SPIE, Caldwell2020RNAAS}). We found that compared to light curves from other community pipelines, the TESS-SPOC light curves are minimally affected by artifacts. However, only a fraction of targets in the FFIs are selected by the TESS-SPOC pipeline. The details of the selection are described in \citet{Caldwell2020RNAAS}, and the impact of this on the discovery potential of SLBs is discussed in Section \ref{ssec:tess_properties}. 

We collected all light curves for objects with TESS magnitudes, $T < 10.5$, which corresponds approximately to an RMS noise of less than 400 ppm at 30-minute cadence (Section \ref{ssec:tess_properties}). Our search is limited to sectors 1 to 69, for which TESS-SPOC light curves were available at the time of writing. We removed any points where the quality flag is $> 0$. This yielded a total of $\sim 900,000$ unique sources and $\sim 2$ million light curves. 

\subsection{Properties of the sample} \label{ssec:tess_properties}

\begin{figure*}
    \centering
    \includegraphics[width=0.99\textwidth]{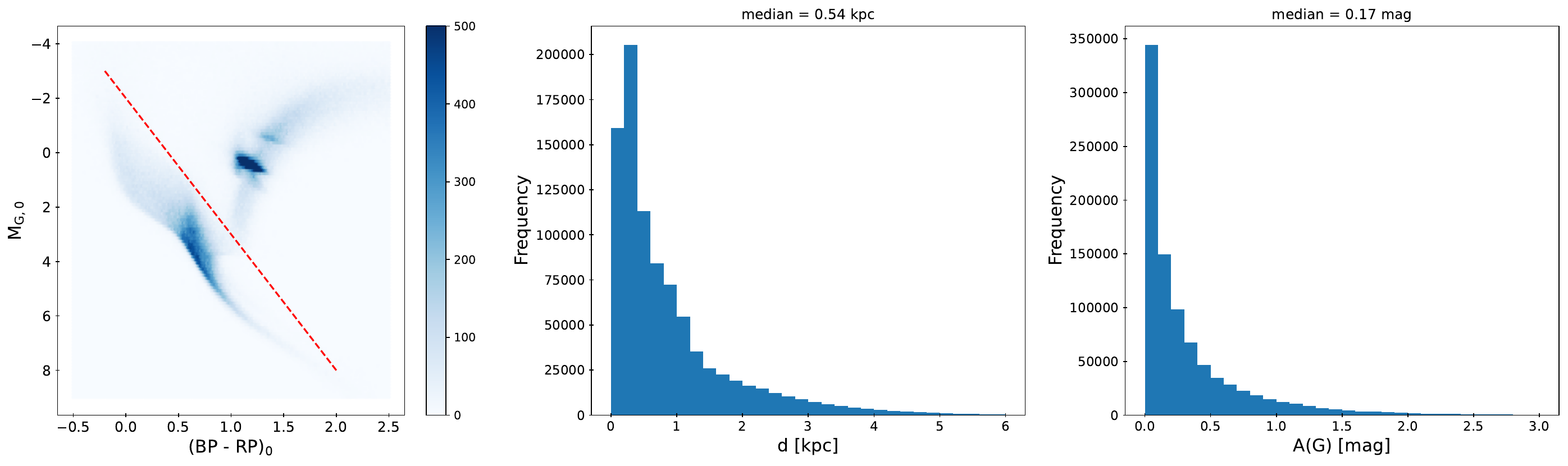}
    \caption{\textit{Left}: Extinction and reddening corrected Gaia CMD of full $T < 10.5$ sample (an upper limit of 500 sources has been set for the color bar so that the MS is clearly seen). Red dashed line shows our empirical cut to select MS stars (Figure \ref{fig:mass_dist}). \textit{Center}: Distance distribution of the same sample. \textit{Right}: Distribution of extinction in the Gaia G band.}
    \label{fig:cmd_dist}
\end{figure*}

\begin{figure*}
    \centering
    \includegraphics[width=0.99\textwidth]{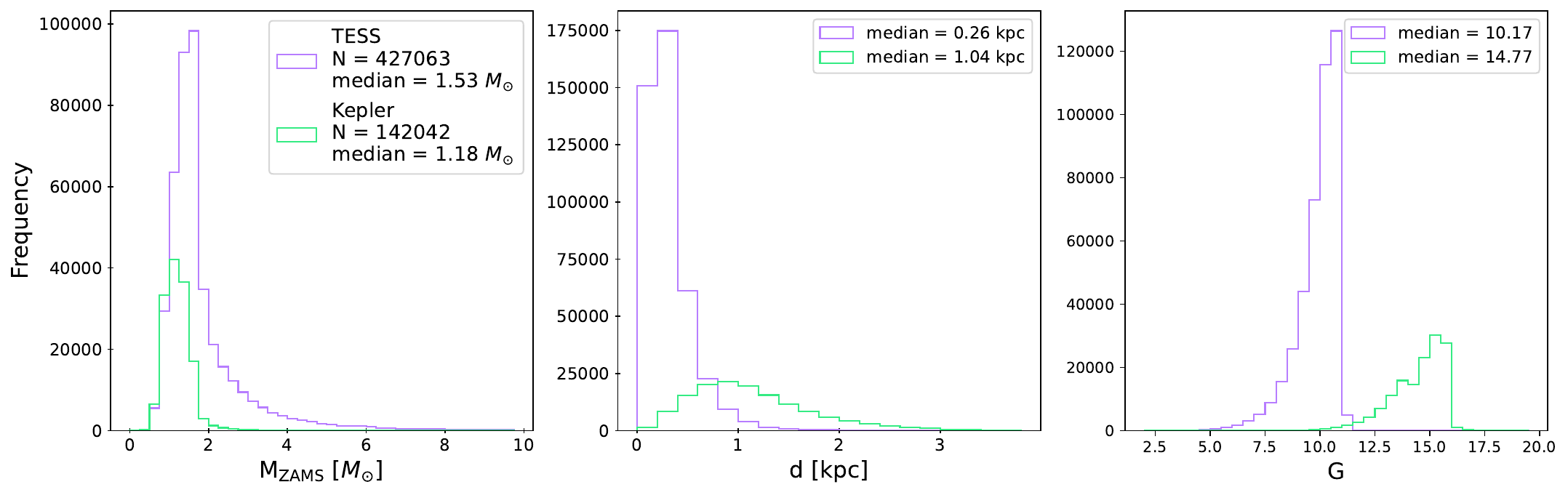}
    \caption{The distribution of masses (\textit{Left}), distances (\textit{Center}), and apparent G band magnitudes (\textit{Right}) for stars on the MS with light curves in our TESS sample and in the Kepler field. TESS probes a wider range of MS star masses and observes nearer and brighter stars than Kepler.}
    \label{fig:mass_dist}
\end{figure*}

\begin{figure*}
    \centering
    \includegraphics[width=0.99\textwidth]{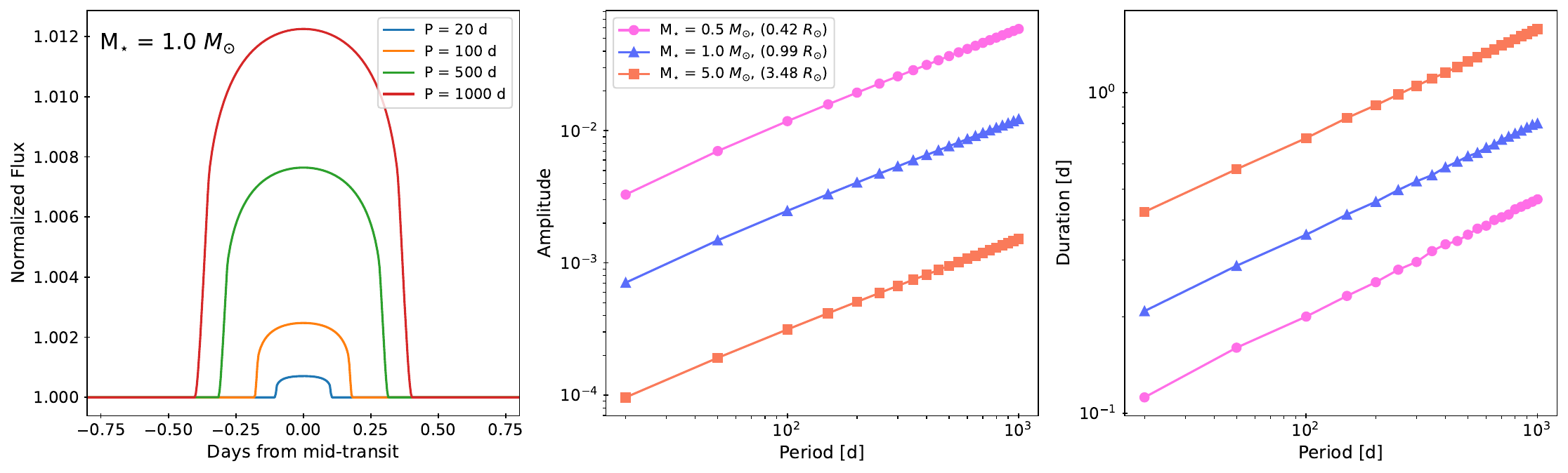}
    \caption{How self-lensing signals depend on the MS star mass, $M_{\star}$, and orbital period, $P$. We assume $M_{\rm WD} = 0.6\,M_{\odot}$ and $i = 90^{\circ}$. \textit{Left}: Light curve models at a fixed mass and several orbital periods from 50 to 1000 days. \textit{Center}: The amplitude of the signal as a function of orbital period for $M_{\star} = 0.5, 1.0,$ and $0.5\,M_{\odot}$. \textit{Right}: The duration of the signal as a function of orbital period for the same masses. From this, we conclude that low-mass stars in long periods have the strongest signals, though these are faint and have the lowest eclipse probabilities.}
    \label{fig:models}
\end{figure*}

We first summarize the bulk properties of our initial TESS sample.

TESS-SPOC provides Gaia DR2 source IDs corresponding to each TESS Input Catalog (TIC) target which we use to match to DR3 source IDs (using the \texttt{dr2\_neighbourhood} catalog). We then obtain reddening/extinction for all objects with parallax $> 0$ using a combination of 3D dust maps (\citealt{Green2019ApJ, Edenhofer2024A&A, Drimmel2003A&A, Marshall2006A&A}, in order of priority) provided by the \texttt{dustmaps} \citep{2018JOSS....3..695M} and \texttt{mwdust} \citep{Bovy2016ApJ} packages. These are then converted to extinctions in Gaia passbands using relations from \citet{Casagrande2018MNRAS}. The left panel of Figure \ref{fig:cmd_dist} shows an extinction-corrected Gaia color-magnitude diagram (CMD). In the middle and right panels, we show distributions of distances and the Gaia G-band extinctions, respectively.

We also estimate masses from absolute magnitudes by interpolating MIST evolutionary tracks \citep{Choi2016ApJ, Dotter2016ApJS}. Here, we assume solar metallicity and assume stars are at zero-age main sequence (ZAMS). As this will be inaccurate for stars that are significantly evolved, we make an empirical cut on the CMD (Figure \ref{fig:cmd_dist}) to select $\sim 430,000$ MS stars. In Figure \ref{fig:mass_dist}, we show the distribution of their inferred masses, as well as their distances and apparent G-band magnitudes. For comparison, we repeated the same process for all Kepler targets and overplot their distributions. We see that TESS-SPOC targets are on average more massive and cover a wider range of masses than Kepler. We also see the advantage of TESS which is that it observes many more stars that are bright and nearby, so any detected SLBs will be easier to follow-up with astrometry and spectroscopy.

\subsection{Dependence of self-lensing signal on stellar and orbital parameters} \label{ssec:LC_models}

We do not explicitly exclude evolved stars in our light curve search. However, these stars have large radii and therefore would not produce self-lensing signals that are strong enough to be detected if the compact object is a WD. In Figure \ref{fig:models}, we show how the shape of the pulse depends on the lensed star's mass and orbital period. Here, we model the pulse as having two components (\citealt{Foreman-Mackey2016AJ}; code available on github\footnote{https://github.com/dfm/transit}): (1) an eclipse of the luminous star by the WD (dimming), and (2) an inverted transit / self-lensing pulse where the radius of the eclipsing star is set to be $\sqrt{2}$ times the Einstein radius of the white dwarf (brightening). The second component is a valid approximation as long as the Einstein radius is much smaller than the background star \citep{Agol2003ApJ}, which is the case for WD + MS binaries that we are interested in. Here, we assume a WD of mass $0.6\,M_{\odot}$ and radius $0.01\,R_{\odot}$, and consider an edge-on orbit. For the lensed star, we assume that it is on the main-sequence and use the \citet{Eker2018MNRAS} mass-radius relation. We take constant quadratic limb-darkening coefficients, $u_1 = 0.4$ and $u_2 = 0.3$ (these have little effect on the amplitude and duration). In the left panel of Figure \ref{fig:models}, we plot the model light curves for different orbital periods at a fixed MS star mass. In the center and right panels, we plot the amplitude and duration against orbital period for a few masses. We see that the amplitude is higher at longer orbital periods; hence, we do not expect to detect SLBs with multiple pulses within a single TESS sector. It also falls rapidly with increasing MS star mass, corresponding to larger radii. Meanwhile, the duration increases with increasing MS star mass and orbital period. 

\begin{figure*}
    \centering
    \includegraphics[width=0.98\textwidth]{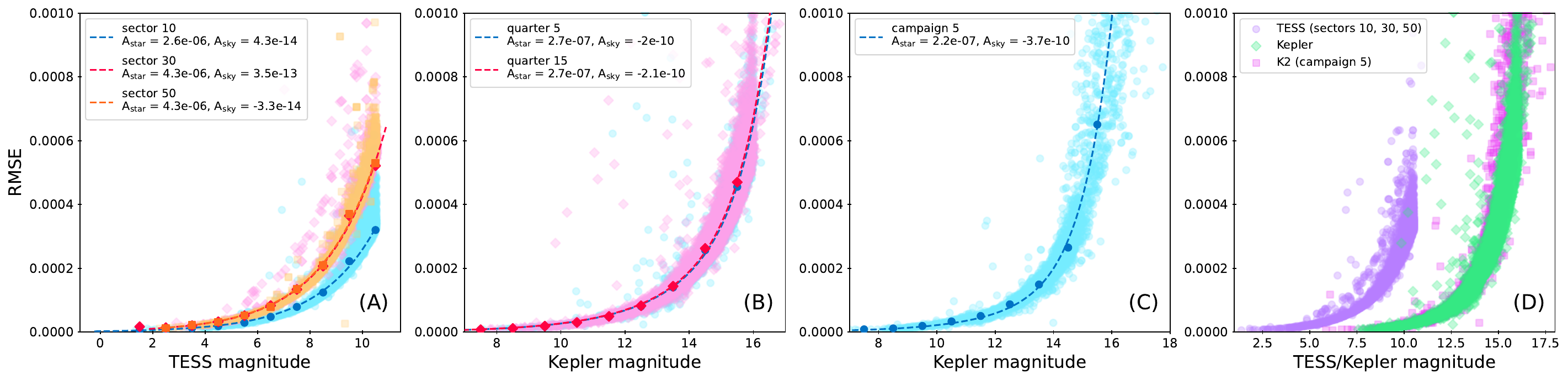}
    \caption{(\textit{A}): The RMS flux error against the TESS magnitude of 5000 randomly selected light curves from sectors 10, 30, and 50. The dashed lines are fits to the median of binned values (darker points) based on a simple model for shot noise associated with the source and background (Equation \ref{eqn:rms_model}). (\textit{B}): Same quantities for Kepler long cadence light curves from quarters 5 and 15. (\textit{C}): Same quantities for K2 long cadence light curves from campaign 5. (\textit{D}): Combination of the previous three plots. The points for TESS sectors 30 and 50 have been scaled down by a factor of $\sqrt{3}$ to account for their 3 times shorter exposure times.}
    \label{fig:rms}
\end{figure*}

\begin{figure*}
    \centering
    \includegraphics[width=0.98\textwidth]{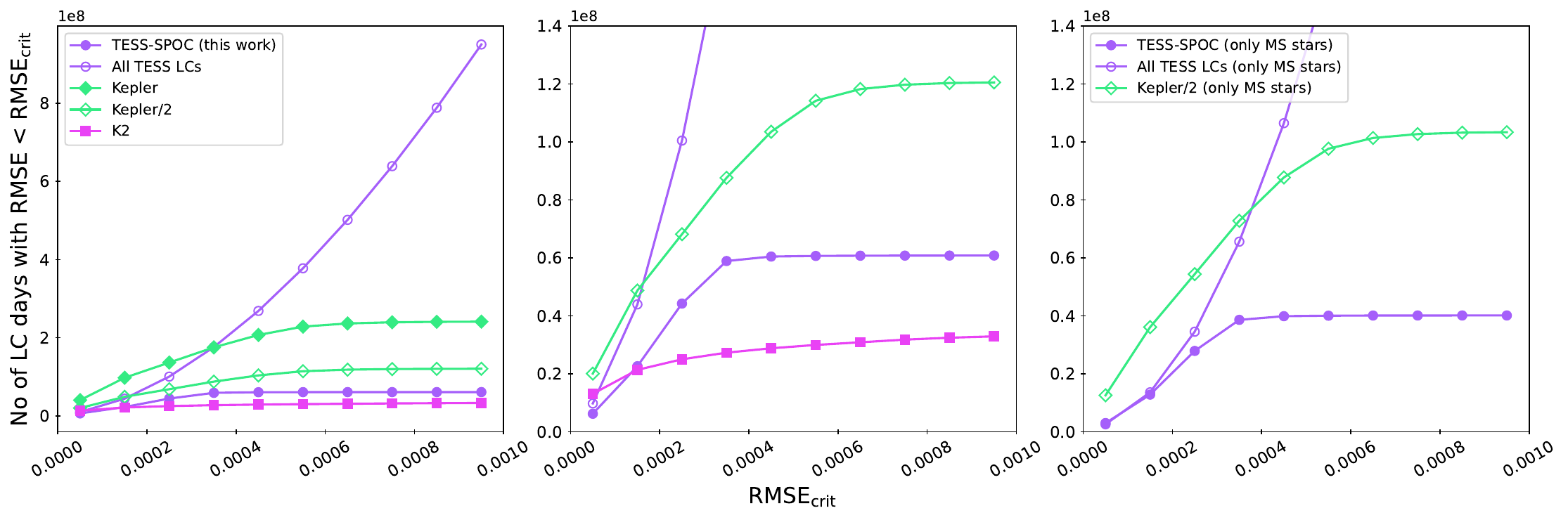}
    \caption{\textit{Left}: Number of light curve days with RMSE below a critical value for TESS (for our sample of TESS-SPOC light curves with $T < 10.5$, as well as all TESS observations), Kepler, and K2. We also plot the Kepler curve rescaled by a factor of 1/2 to account for its longer than needed baseline to observe a single transit of a typical SLB with orbital period $\sim 100 - 1000\,$d. RMSE have been standardized to 30 minute exposures. \textit{Center}: Similar plot, but zoomed in for easier comparison between our sample and Kepler. \textit{Right}: Curves for TESS and Kepler only including observations of MS stars, since self-lensing pulses due to WDs orbiting giants will have amplitudes too low to be detected.}
    \label{fig:lc_days}
\end{figure*}

\subsection{SLB discovery potentials} \label{ssec:discovery_potential}

Here, we estimate the relative SLB discovery potentials of TESS and Kepler. For reference, we also make some comparisons to the extended Kepler mission, K2 \citep{Howell2014PASP}. 
 
In the leftmost panel of Figure \ref{fig:rms} (panel A), we plot the root mean squared flux errors (RMSE) against the TESS magnitude. We bin these and take their median values, which we fit with a simple relation accounting for the shot noise from the star and sky background \citep{Ricker2014SPIE}: 
\begin{equation}
    \mbox{RMSE} =\sqrt{\frac{A_{\rm star}^2}{10^{-T/2.5}} + \frac{A_{\rm sky}^2}{10^{-2T/2.5}}}
\end{equation} \label{eqn:rms_model}
The best-fit values of the coefficients are listed in the figure. We see that shot noise from the star dominates over background noise over the magnitude range of interest for our sample (while background noise dominates at $T \gtrsim 13$), and thus the RMSE increases at fainter magnitudes such that the signal-to-noise ratio (SNR) scales with the square root of the flux. Compared to sector 10, the errors for sectors 30 and 50 are larger by a factor of $\sim \sqrt{3}$ across all magnitudes. This is a consequence of the FFI cadence being increased from 30 to 10 minutes starting from sector 27 (then 200 seconds from sector 56).

In panel B, we show the same quantities but for Kepler long cadence (30 minute) light curves from quarters 5 and 15. In panel C, we show one campaign of K2. Finally, in panel D, we compare all missions on the same axis. Note that here, the RMSE of sectors 30 and 50 for TESS have been divided by $\sqrt{3}$ to match the 30 minute cadence of the other data. While the TESS and Kepler passbands not identical, Kepler's advantage of having a $\sim 100$ times larger mirror is clear: Kepler light curves have similar SNR to TESS light curves that are 5 magnitudes brighter.

To quantify how much more likely a transit is to be observed in different missions, in Figure \ref{fig:lc_days}, we plot the cumulative number of light curve days with RMSE below some critical value. This is simply the cumulative number distribution of RMSE for all light curves, multiplied by the duration of observation. We plot this for our sample of TESS-SPOC light curves ($T < 10.5$), each contributing $25$ days (the approximate number of usable days per sector, given gaps in the data). Once again, RMSE for light curves starting from sector 27 (56) were rescaled by a factor of $1/\sqrt{3}$ ($1/\sqrt{9}$). 

However, this is not illustrative of the maximum potential of TESS because only a subset of its observations are processed by TESS-SPOC \citep[For a description of the selection, see ][]{Caldwell2020RNAAS}. Therefore, we add another curve for all observations (between sectors 1 - 69) of all sources from the TIC. We check which sectors each TIC source with $T < 13$ was observed in using \texttt{tess-point} \citep{2020ascl.soft03001B}, precomputing the observed sectors on 49152 uniformly spaced points on the sky to save processing time \citep{Zonca2019, 2005ApJ...622..759G}. We then estimate RMSE from apparent magnitudes using fits as shown in Figure \ref{fig:rms}. The number of unique sources with $T < 13$ (corresponding to RMSE $\lesssim 0.001$, the typical amplitude of a self-lensing pulse due to a WD companion) is over $12$ million. Compared to our sample, the number of light curve days with RMSE $< 0.0004$ is larger by a factor of $\sim 3$. While we only analyze TESS-SPOC light curves in this search, this tells us that including light curves from all TESS observations may largely increase the number of detected SLBs. 
 
Since Kepler observed the same field continuously over its 4-year mission and RMSE as a function of apparent magnitude remains roughly constant across quarters (Figure \ref{fig:rms}), we assume 4 years of observations with the RMSE reported in a single quarter. Even with there being over $5$ times more unique sources in our sample of TESS-SPOC light curves compared to the total number observed by Kepler, Kepler produced almost 4 times more light curve days with RMSE $< 0.001$. This is primarily a result of Kepler's larger aperture. However, to detect a non-repeating self-lensing signal, it is only necessary to observe an SLB for at most the length of its orbital period, which we expect to be in the range of $\sim 100 - 1000\,$d for WD lenses. This means that beyond a $\sim 2\,$year long baseline, the probability of observing a transit for a given star does not increase significantly. Thus, for a fairer comparison, we also plot Kepler's curve rescaled by a factor of $1/2$. Furthermore, since self-lensing is stronger for MS companions compared to giants (Section \ref{ssec:LC_models}), in the rightmost panel, we plot the curves for TESS and Kepler including only MS stars (based on their positions on the CMD; Figure \ref{fig:cmd_dist}). We see that TESS observes a larger fraction of giants, and overall, the SLB discovery space for Kepler light curves is $\sim$2.5 times larger than that of our TESS-SPOC sample. Meanwhile, considering all TESS observations increases the number of observed sources with RMSE $< 0.001$ to $\sim70$ times more than Kepler, resulting in almost $4$ times more light curve days, more than compensating for the shorter baselines. It is worth noting however that for MS stars, Kepler still outperforms TESS below an RMSE of $\sim 0.0004$. 

We also add approximate curves for K2. We used the fit plotted in Figure \ref{fig:rms} to estimate RMSE of light curves from all campaigns given their Kepler magnitudes. We therefore only include sources from the EPIC catalog that have reported magnitudes \citep{Huber2016ApJS}, which is $\sim 60\%$ of all sources, so the curve was rescaled by a factor of 1.5 to account for this. We take 80 days to be the length of observation in one campaign. We see that K2 does slightly worse than the TESS-SPOC sample and significantly worse than Kepler. This is because while the total duration of the K2 mission was similar to the original Kepler mission, the number of stars per field was $\sim 5$ times lower on average, each with a shorter baseline \citep{Howell2014PASP}. Based on this result, we chose not to expand our search to include K2 data.  

\section{Light curve search} \label{sec:main_search}

%If not already discussed earlier, should explain that previous searches with Kepler were easier because they could search for repeating transits, but we can't do that because TESS sectors are short

Unlike previous searches for SLBs with Kepler, we do not attempt to find multiple transits because the shorter duration of TESS sectors makes this unlikely. Instead, we closely followed the approach taken by \citet{Foreman-Mackey2016AJ} who searched Kepler light curves for long-period planets with only a single detected transit. We made minor modifications to their publicly available code\footnote{https://github.com/dfm/peerless} to apply in our search for non-repeating self-lensing signals. We refer readers to their paper and code for details, but provide a summary here. 

Firstly, bad data points with missing fluxes were removed and the light curves were split into continuous chunks with gaps no longer than 0.2 days. This reduces the number of false positive signals centered on breaks in observations during data downlink. Light curves with 30-minute and 10-minute cadence were analyzed at their native cadence, but those with 200-second cadence were rebinned to 10-minute cadence to reduce computation time. The light curves were then de-trended with a running median calculated in a 2 day window and the signal-to-noise (S/N) was computed using a box-shaped pulse model. Candidate signals with S/N $> 10$ were identified and proceeded to the next step in the analysis. In typical sectors, 1-2\% of all light curves exhibited at least one such candidate event.

Next, we looked for pulses whose shape is better described by a self-lensing model than by several other possible models. This is done by carrying out Gaussian Process (GP) fitting of the models to each pulse and measuring the goodness of fit with the Bayesian information criterion (BIC). For this, we used the \texttt{george} python package \citep{2015ITPAM..38..252A}. The same models as those in \citet{Foreman-Mackey2016AJ} were used, which are (1) a transit (in our case, self-lensing) model with realistic limb-darkening (e.g. left panel of Figure \ref{fig:models}), (2) a pure GP model, which is meant to model stellar variability, (3) a single outlier model, which excludes one bad data point, and (4) a box-shaped model. We exclude the ``step" model used by \citet{Foreman-Mackey2016AJ} since we found that such signals did not significantly contaminate the resulting sample. We refer readers to their paper (Appendix A) for mathematical descriptions of the models. The box and transit models were inverted to model positive pulses as opposed to negative dips. At this stage, we did not fit for stellar and/or orbital parameters. Instead, we fit for the height, duration, and central time of each pulse. Pulses for which the inverted transit model yielded a larger BIC than any other model were retained. This narrowed down candidates to $\sim 50$ per sector. We do not attempt in this work to provide physical explanations for all the other signals, but we discuss some of the most common false-positives below.  

\begin{figure*}
    \centering
    \includegraphics[width=0.95\textwidth]{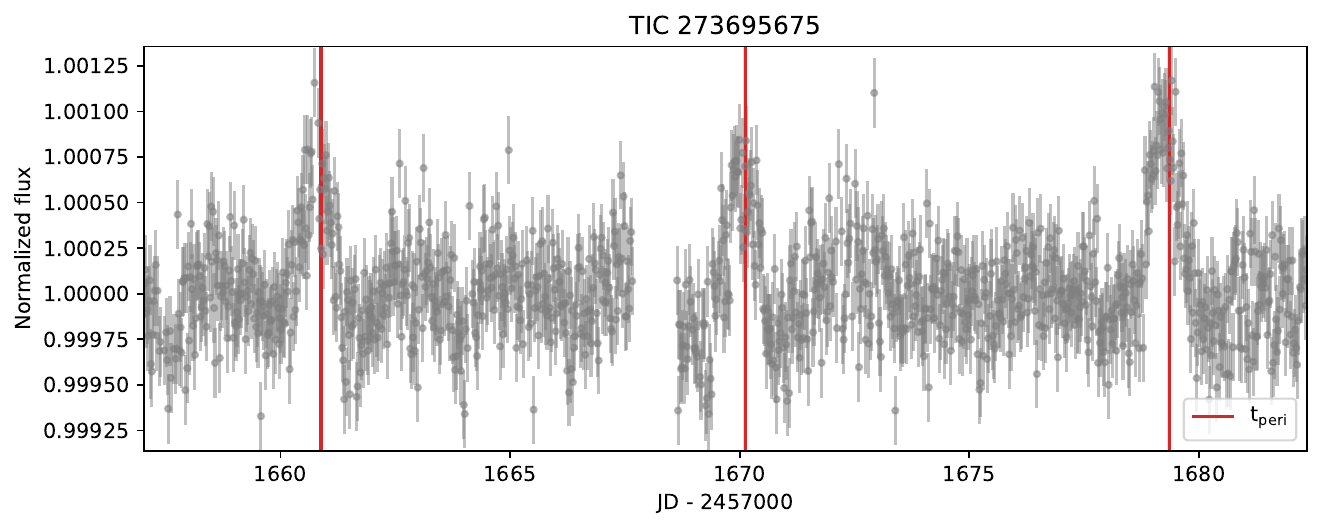}
    \caption{The de-trended light curve of a star showing heartbeat-like signals. This system has a double-lined spectroscopic orbital solution in Gaia DR3 (ID 4618236260167319296). As expected for heartbeat systems, it has a moderate eccentricity of 0.37 and the pulses occur at the periastron times predicted by the orbital solution, marked with red vertical lines. }
    \label{fig:heartbeat}
\end{figure*}

\begin{figure*}
    \centering
    \includegraphics[width=0.95\textwidth]{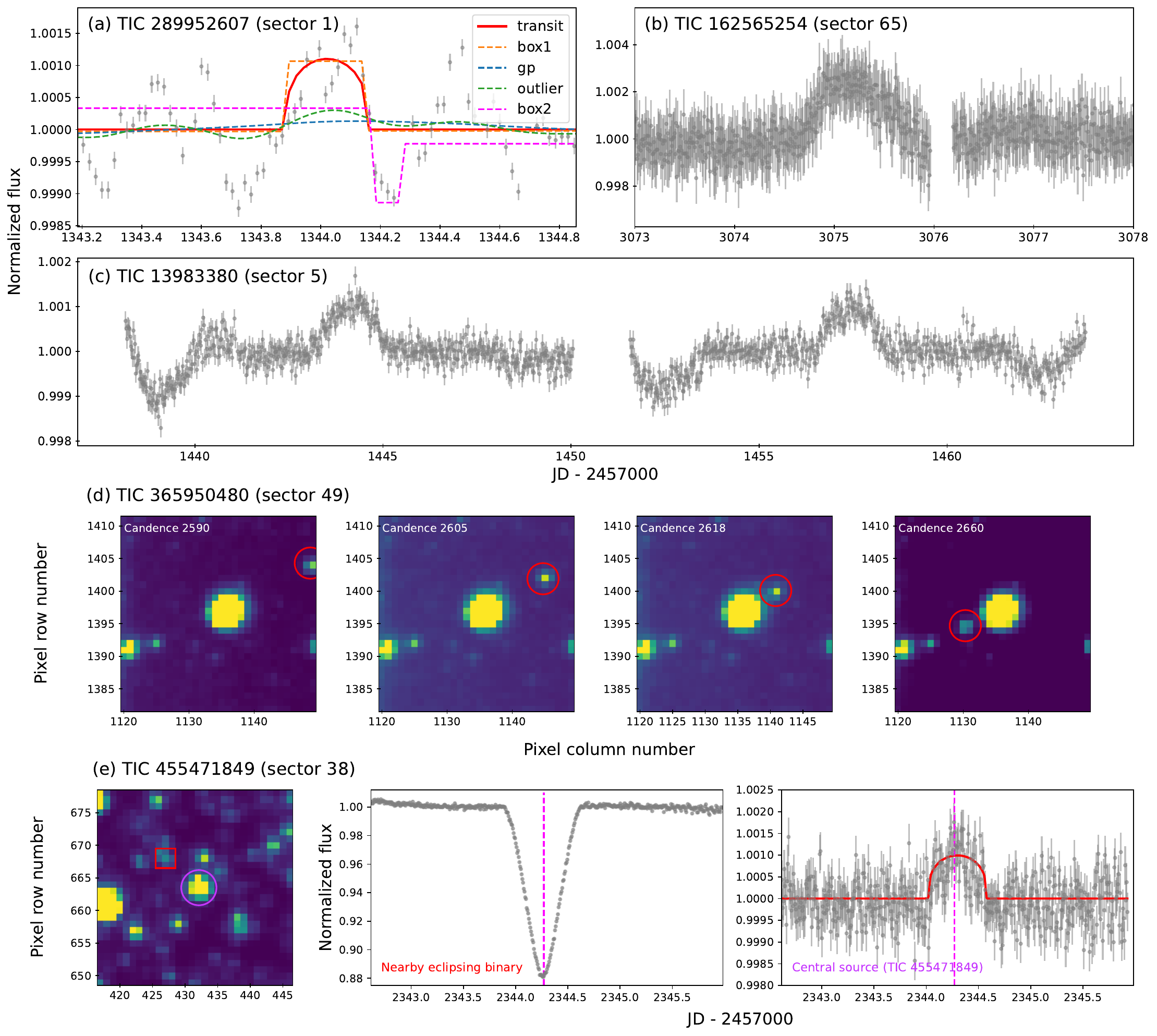}
    \caption{Examples of common false positives in our sample (other than heartbeats; Figure \ref{fig:heartbeat}): (a) Pulses that are not well fit by any model -- here, we show five different models that have been fit to a pulse, none of which are a good representation of the data but where a self-lensing model had the highest BIC, (b) Artifacts occurring at the edge of data gaps, (c) Repeating signals with unclear origins, but with periods too short to be self-lensing, (d) Asteroid crossing events, as seen in FFIs taken around the mid-pulse time of the central source. (e) Over-subtraction of background eclipses. The light curve in the middle shows dimming of a nearby star (selected pixels boxed in red in the FFI) to the central source which occurs at the same time as the detected pulse.}
    \label{fig:false_positives}
\end{figure*}

\subsection{Quality cuts}

To cut down on the number of candidates further, we selected only pulses with best-fit durations greater than $0.2\,$d, and amplitudes smaller than 0.04 that occurred in light curves of sources with absolute extinction-corrected Gaia G magnitudes $M_G,0 > 0.4$. These are conservative cuts based on models of expected self-lensing pulses over a range of WD masses, companion star radii, and orbital periods. For example, much stronger signals (i.e. larger amplitudes) would require lens masses significantly above the $1.4\,M_{\odot}$ limit for a WD or long orbital periods of $\gtrsim 1000\,$d, where the transit probability is low. The lower limit on the G magnitude removes most giants and MS stars with mass $\gtrsim 3 M_{\odot}$, where self-lensing pulses are weak (Section \ref{ssec:LC_models}) and which we find present more false positives (see discussion of heartbeat stars below). We also removed all sources with a total RV variability (\texttt{rv\_amplitude\_robust}) measured by Gaia to be $< 5\,\mbox{km s}^{-1}$ (see discussion below regarding Figure \ref{fig:rv_amp}; sources with no measured value was kept). While the five Kepler SLBs are too faint to have measured \texttt{rv\_amplitude\_robust}, they would have satisfied all these cuts if they were brighter. After this step, about $10$ objects remained per sector, leaving a total of $\sim 800$ candidate signals across all sectors. 

\subsection{Visual inspection \& false-positives}

We then inspected every remaining candidate by eye. The most common false positives appear to be heartbeat stars, with repeating pulses on timescales of days to weeks \citep[e.g.][]{Welsh2011ApJS, Fuller2017MNRAS}. While we did not follow up such systems and thus cannot confirm their classification, there are several with radial velocity orbital solutions from Gaia DR3 \citep{GaiaCollaboration2023A&A} where the times of pulses aligned with the expected periastron times, strongly indicating that these are in fact heartbeat stars. These also typically show a dip in the flux before and after the peak. Figure \ref{fig:heartbeat} shows an example of one such object. % ex. TIC273695675 - Sector 13, TIC364394197 - Sector 8
(For more on heartbeat systems from TESS, see recent work by \citealt{Solanki2024arXiv}.) Examples of other common contaminants are shown in Figure \ref{fig:false_positives}. These are signals badly fit by all of the models (panel a), increasing flux at the edges of data gaps due to poor de-trending (b), repeating signals at periods too short for self-lensing and which have shapes distinct from typical heartbeat stars (c), and asteroid crossing events (also found to be significant contaminants by \citet{Sorabella2023ApJ}; a detailed description of one such false positive can be found in Appendix \ref{appendix:TIC_55276301}) (d). To identify and remove possible asteroid crossing events, we inspected FFIs \citep{Brasseur2019ascl} $\sim 1$ day around the central time of the peak to look for faint sources moving across the camera and traveling over the source. FFIs also revealed several cases of likely background eclipsing binaries, seen as a dimming in a source located close to the target, which can produce a positive pulse in surrounding pixels due to background over-subtraction (e).

\subsection{Gold sample}

Finally, we selected a `gold sample' of the most promising candidates for multi-epoch spectroscopic follow-up. Where available, we used the Gaia DR3 orbital solutions to assess whether sources are consistent with hosting WD companions. We removed systems with low inclinations, short periods, and double-lined spectroscopic (``SB2") solutions. In the absence of orbital solutions, we prioritized sources with Renormalized Unit Weight Error (RUWE) $> 1.4$ and \texttt{rv\_amplitude\_robust} $> 10\, \mbox{km/s}$ which can be suggestive of binarity at the expected orbital periods. RUWE is a measure of the how well a single-star model fits the observed astrometry, where values close to 1 indicate a good fit \citep{LL:LL-124}. 

\begin{figure*}
    \centering
    \includegraphics[width=0.8\textwidth]{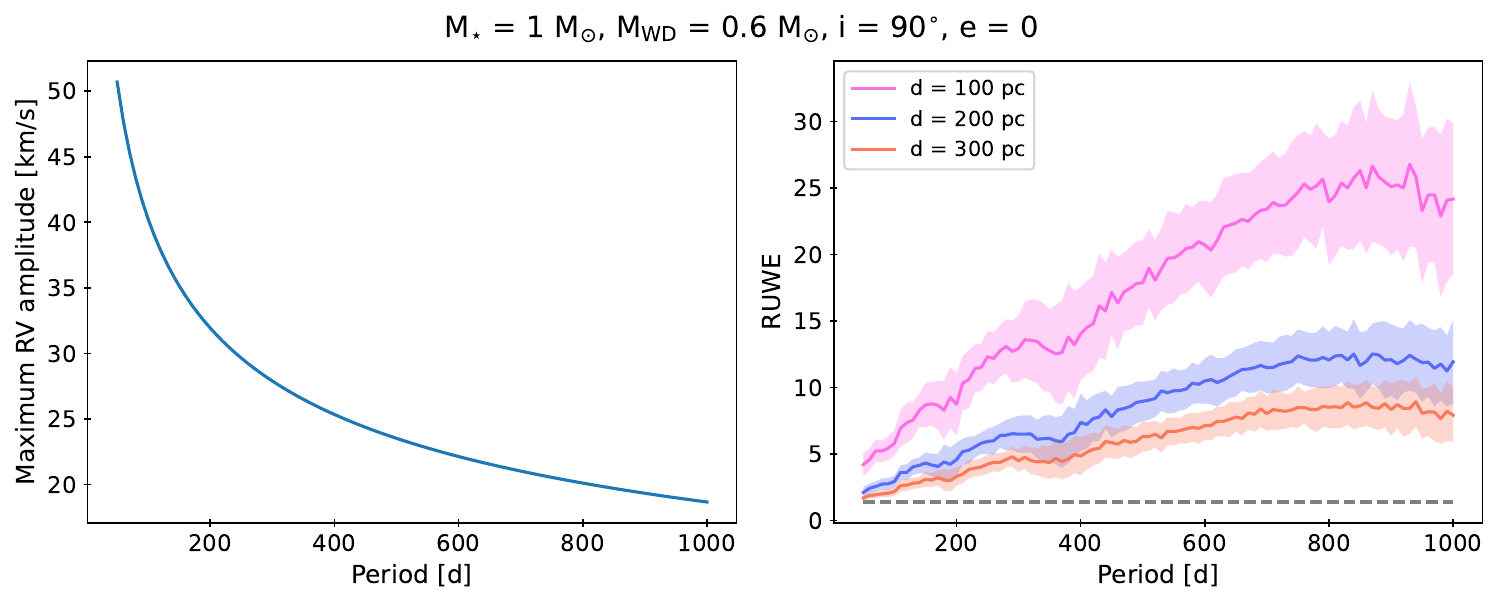}
    \caption{\textit{Left}: Total RV variability amplitude as a function of orbital period for an edge-on binary consisting of a $0.6\,M_{\odot}$ WD and $1.0\,M_{\odot}$ star, with zero eccentricity. We see that this value is $\sim 18\,$km/s at a period of 1000 days. \textit{Right}: RUWE as a function of orbital period at different distances for the same system. The shaded region shows the $1-\sigma$ spread in the RUWE for 100 randomly chosen coordinates and orientations. We see that at a distance of $\sim 100\,$pc and period of 50 days, the predicted RUWE is $\sim 1.4$ (dashed gray line). On the basis of these plots, we prioritize candidates with Gaia \texttt{rv\_amplitude\_robust} $>$ 10 km/s and RUWE $> 1.4$.}
    \label{fig:rv_amp} 
\end{figure*}

In Figure \ref{fig:rv_amp}, we plot the maximum RV variability amplitude (i.e. two times the semi-amplitude) and predicted RUWE as a function of orbital period for a $1.0\,M_{\odot}$ star hosting a $0.6\,M_{\odot}$ companion. We assume an edge-on orbit and zero eccentricity. We predicted RUWE using the \texttt{gaiamock} package, which generates mock Gaia epoch astrometry of astrometric binaries following the Gaia scanning law and fits the data with a single-star model (\citealt{El-Badry2024OJAp}; code available on GitHub\footnote{https://github.com/kareemelbadry/gaiamock}). We generate 100 realizations of the same binary with the specified parameters at random coordinates and orientations. We shade the region $1\sigma$ away from the median of the distribution at each period and distance. For periods below 1000 days, we expect RV amplitudes $\gtrsim 20\,$km/s and above 50 days, RUWE $\gtrsim 1.4$. Note that we did not place a stricter cut on \texttt{rv\_amplitude\_robust} as the exact value of this depends on the fraction of an orbit that is observed (as well as the exact orbital and stellar parameters). The results of our RV follow-up will be presented in future work. 

In Table \ref{tab:candidates_list} of Appendix \ref{appendix:candidate_tables}, we summarize properties of the most promising targets after visual inspection. We include qualitative notes of their key properties, which determine their order of priority for follow-up. We also plot the detected signals and best-fit model of the 12 most promising candidates in Figure \ref{fig:best_candidates}. 

\subsection{Focused search in the Gaia NSS catalog} \label{ssec:nss_search}

We also performed a search isolating objects in the Gaia DR3 non-single stars catalog. In many cases, a spectroscopic or astrometric orbital solution can immediately rule out candidates whose mass functions or inclinations are inconsistent with being SLBs. These solutions also provide constraints on the periastron time, period, and eccentricity so it is possible to check whether a candidate signal occurs during the expected time of eclipse. Furthermore, fewer observations are needed to constrain the radial velocity curves, and thus orbits, of these systems. 

Here, we removed the requirement of a best-fit transit model as well as any cuts based on the amplitude and duration of the pulse. Instead, we allowed periods between 50 and 1000 days which is in the expected range for self-lensing to be detectable, and \texttt{goodness\_of\_fit} $< 4$ (large values indicate a poor fit to the astrometric model). In the case of astrometric binaries, we only kept objects with inclinations within 5$^{\circ}$ of edge-on (otherwise, we would not expect an eclipse). Once again, we excluded systems with SB2 solutions. 

Many astrometric binaries were ruled out based on their inclinations, and others were identified to be asteroid crossing events from their FFIs. We include one additional candidate from this search in Table \ref{tab:candidates_list}. A few of our candidates have SB1 solutions. While SB1s do not have constraints on the inclination, we can calculate a lower limit on the companion mass which occurs for an edge-on orbit. In the case that the system is indeed a SLB, this lower limit must be close to the true value. We removed several systems in eccentric orbits, with $e \gtrsim 0.3$. While this does not rule them out as SLBs, they are more likely to be heartbeat systems and possibly in tension with being post-interaction WD + MS binaries which have been tidally circularized (though a few such eccentric binaries are present in the Gaia sample; \citealt{Shahaf2024MNRAS, Yamaguchi2024PASP_2}).  

%In Table \ref{tab:candidates_list}, we include one good candidate from this search. Many astrometric binaries were ruled out based on their inclinations. While SB1s do not have constraints on the inclination, we can calculate a lower limit on the companion mass which occurs for an edge-on orbit. In the case that the system is indeed a SLB, this lower limit must be close to the true value so we prioritize systems where it is reasonable for a WD companion. The orbital periods vary from $\sim 50$ to $1000$ days which is in the expected range for self-lensing to be detectable. Several systems are in eccentric orbits, with $e \gtrsim 0.3$. While this does not rule them out as SLBs, they are more likely to be heartbeat systems and possibly in tension with being post-interaction WD + MS binaries which have been tidally circularized (though a few such eccentric binaries are present in the Gaia sample; \citealt{Shahaf2024MNRAS, Yamaguchi2024PASP_2}).  

\section{Injection and Recovery tests} \label{sec:injection_and_recovery}

To test the detection efficiency of our search, we carry out two injection and recovery tests. 

\subsection{A fixed self-lensing pulse} \label{ssec:injection_and_recovery_1}

\begin{figure}
    \centering
    \includegraphics[width=0.98\columnwidth]{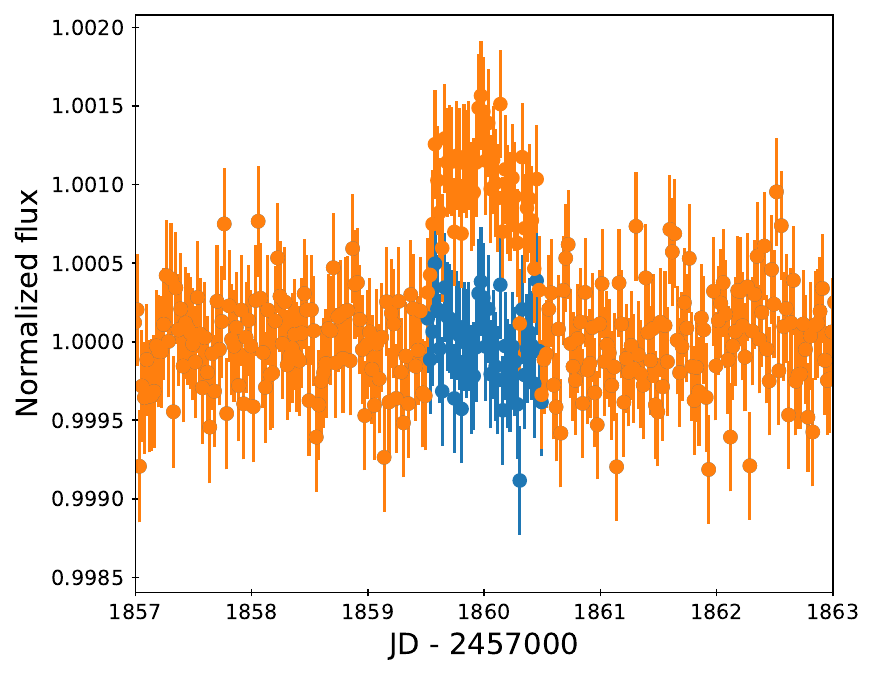}
    \caption{An example of a light curve before and after an artificial self-lensing pulse has been injected.}
    \label{fig:injected_signal}
\end{figure}

For the first test, we inject a fixed self-lensing pulse into 5000 randomly selected TESS-SPOC light curves in a few sectors. For each light curve, we choose a central time at random and inject an inverted transit model with an amplitude of $\sim 0.001$ and duration of a day, comparable to the Kepler SLBs of \citet{Kawahara2018AJ}. Figure \ref{fig:injected_signal} shows a portion of a light curve before and after a signal has been added. We then run these light curves through our algorithm, tracking injected signals which are simply ``detected" (i.e. make it past our first S/N cut) and amongst those, signals which are best fit with the inverted transit model (thus making it into our initial list of candidates, before further cuts based on stellar parameters and manual inspection). 

\begin{figure*}
    \centering
    \includegraphics[width=0.95\textwidth]{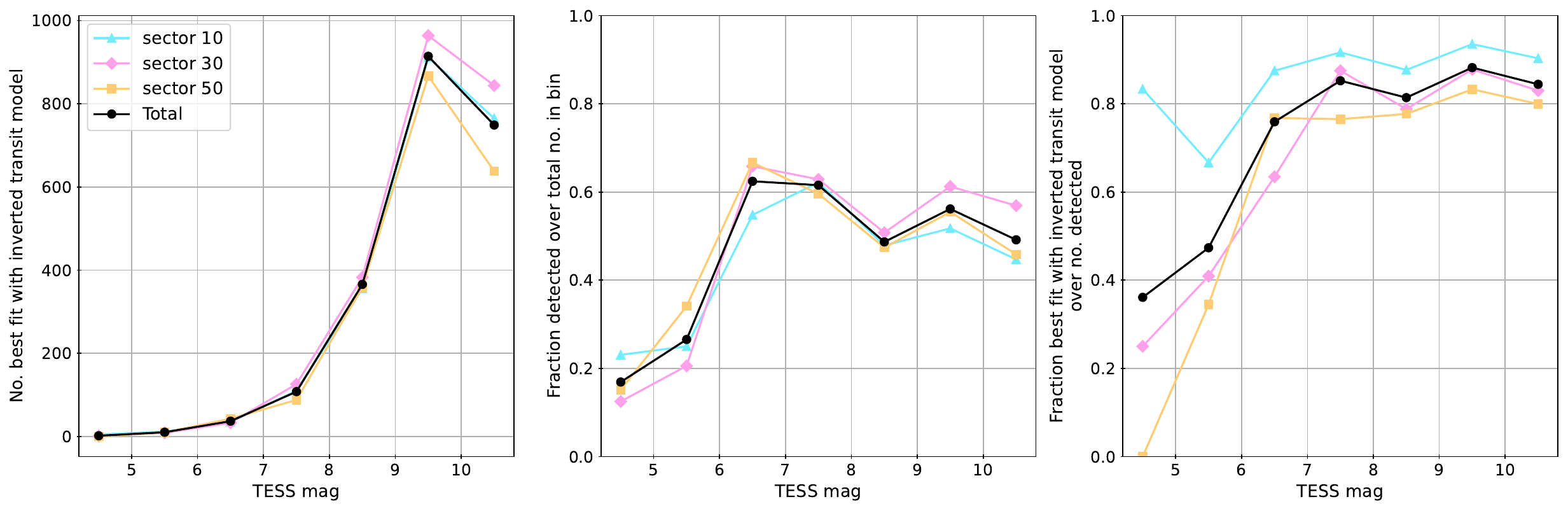}
    \caption{Recovery rate of injected signals in sectors 10, 30, and 50. The black line is the combined result. \textit{Left}: The apparent magnitude distribution of the 5000 objects whose light curves were used for this test.  \textit{Center}: Fraction of the total number of injected signals which were detected with S/N $>$ 10. \textit{Right}: Fraction of the detected signals which were best fit with the inverted transit model.}
    \label{fig:injected_signal_test}
\end{figure*}

\begin{figure*}
    \centering
    \includegraphics[width=0.95\textwidth]{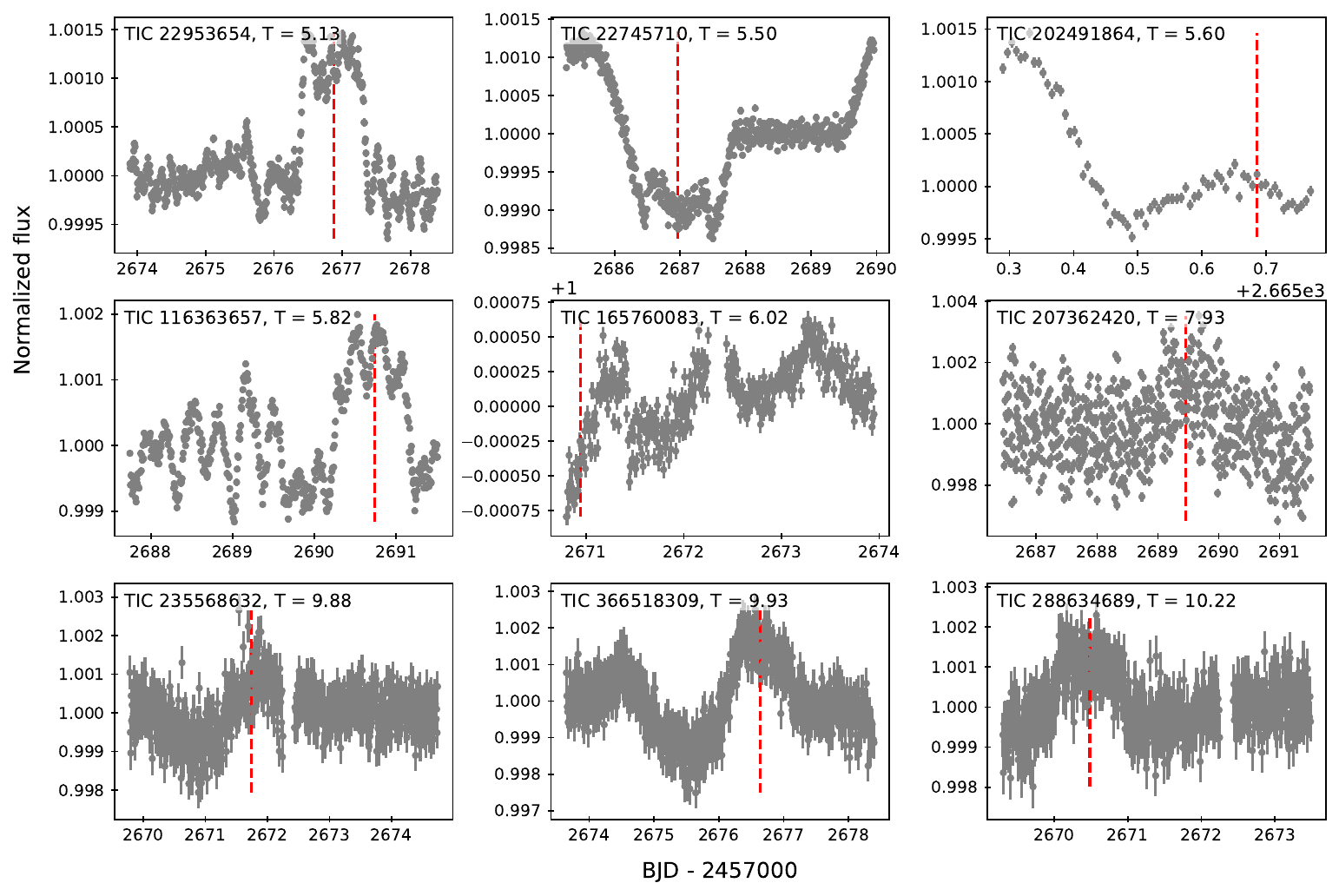}
    \caption{Example light curves of sources in sector 50 where the artificial self-lensing signals were not detected. Red dashed lines indicate the central time where the signals were injected. We see that the brightest stars have intrinsic variability which results in the injected self-lensing pulse having insufficient SNR to be detected.}
    \label{fig:signals_not_detected}
\end{figure*}

In the leftmost panel of Figure \ref{fig:injected_signal_test}, we plot the number distribution of the apparent TESS magnitudes of the 5000 objects whose light curves were randomly chosen for this test. As might be expected, there are significantly more faint objects with $T \gtrsim 8$ than there are brighter ones. In the middle, we plot the fraction of objects whose injected signals were detected with S/N $>$ 10 as a function of apparent magnitude. Between T $\sim$ 6 - 10.5, this fraction remains roughly constant (Figure \ref{fig:rms}), averaging at about 60\%. However, for stars brighter than $T \sim 6$ (which represent only $\sim 1.4$\% of the full sample), the recovery rate falls. We plot the light curves of several objects in Figure \ref{fig:signals_not_detected} where the injected signals were not detected. We find that overall, most of the brightest stars -- which are a mix of giants and massive stars -- display intrinsic variability, preventing the identification of injected self-lensing pulses with high S/N. Finally, in the right panel of Figure \ref{fig:injected_signal_test}, we plot the fraction of the detected signals which were best fit by the inverted transit model in each magnitude bin. We see that this fraction is high, ranging from $\sim$ 80 to 90\% for $T \gtrsim 7$. Once again, the fraction falls for the brightest objects. Overall, this test tells us that about $50\%$ of all self-lensing binaries with amplitudes comparable to the known systems will make it into our final list of candidates. 

\subsection{Simulating the WD + MS binary population} \label{ssec:injection_and_recovery_2}

\begin{figure*}
    \centering
    \includegraphics[width=0.99\textwidth]{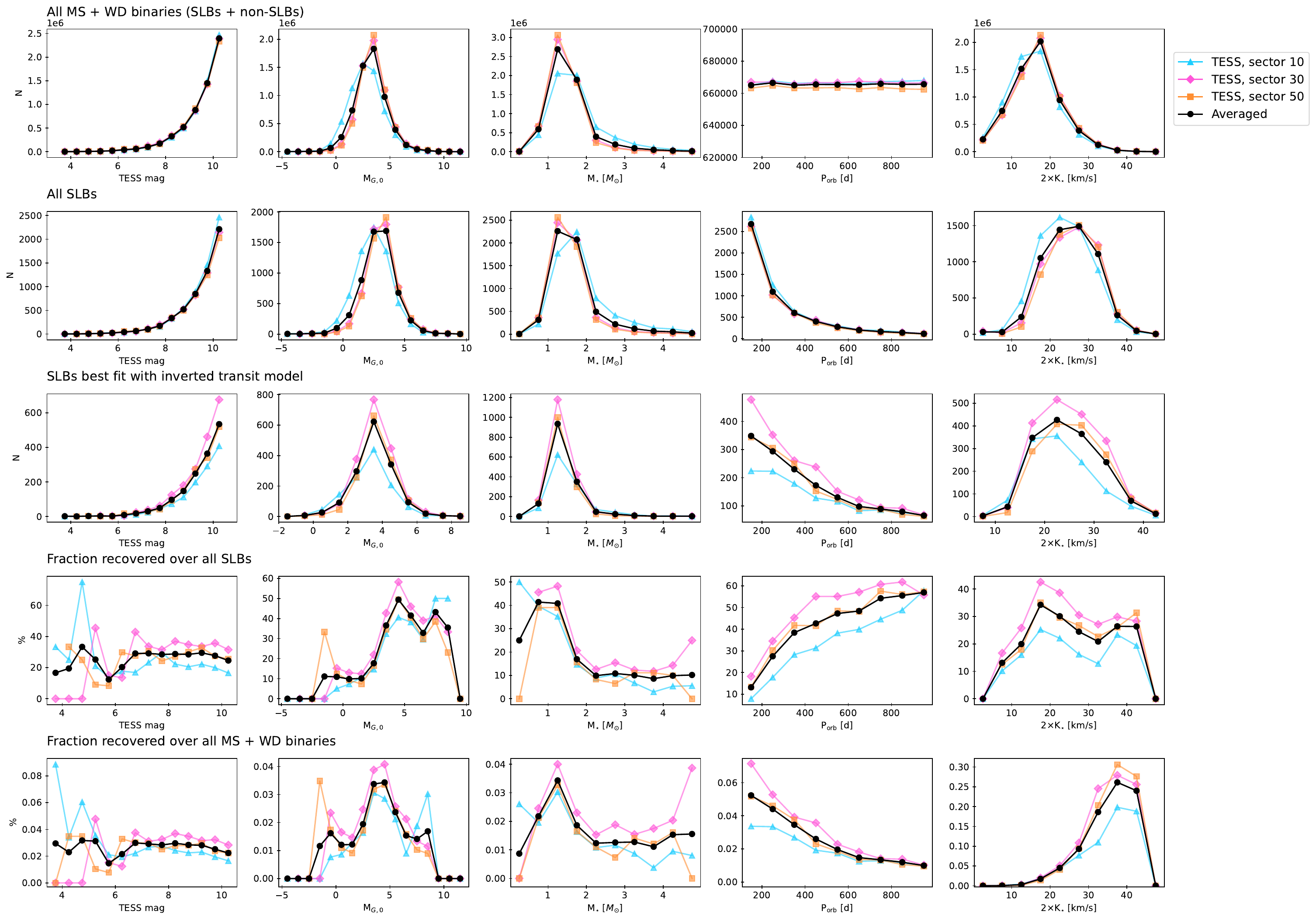}
    \caption{The results of the injection and recovery test using light curves from several different sectors to simulate a population of WD + MS binaries with orbital periods between 100 and 1000 days. Each row plots the same value, binned by different stellar and orbital parameters. From top to bottom, this is: (1) the number of all simulated WD + MS binaries, (2) the number of binaries showing self-lensing, (3) the number of self-lensing signals recovered by our algorithm, (4) the fraction of recovered signals over all SLBs, and (5) the fraction of recovered signals over all WD + MS binaries. From left to right, these are binned by: (1) TESS magnitude, (2) absolute G band magnitude, (3) stellar mass, (4) orbital period, and (5) RV amplitude. In black, we plot the median value across the four sectors.}
    \label{fig:injected_signal_test_2}
\end{figure*}

Ultimately, we would like to infer the occurrence rate of AU-scale WD + MS binaries, based both on eventually confirmed SLBs from our search and on the Kepler SLB sample (Section \ref{sec:rates}). To measure the selection function of our search and make this modeling possible, we generate a population of WD + MS binaries, identify those that are self-lensing, inject artificial signals into the lightcurves and run them through our pipeline. We detail the steps below: 

\begin{enumerate}
    \item Select MS stars ($N \sim 430,000$; Figure \ref{fig:cmd_dist}) and use their masses, $M_{\rm star}$, estimated from their absolute G magnitudes (Figure \ref{fig:mass_dist}). We estimate their radii, $R_{\rm star}$, using the \citet{Eker2018MNRAS} mass-radius relation.
    \item For each of these simulated WD+MS binaries, randomly select an orbital period, $P_{\rm orb}$, between 100 and 1000 days, and cosine of the inclination, cos($i$), between 0 and 1, in both cases assuming a uniform distribution. We fix the WD mass, $M_{\rm WD} = 0.6\,M_{\odot}$, eccentricity, $e = 0$, and quadratic limb-darkening coefficients, $u_1 = 0.4, u_2 = 0.3$ (roughly corresponding to the values for a solar-type star).  
    \item Given these orbital properties, calculate the transit duration, $t_{\rm dur}$ using equation 3 of \citet{Seager2003ApJ} (appropriate for circular orbits). This equation is undefined in the case that there is no transit (i.e. when $a\mbox{cos}(i)/R_{\star} > 1$, where $a$ is the semi-major axis). The probability of observing the transit during a single TESS sector scales as $\sim (t_{\rm dur} + 2 \cdot 25\,\mbox{d})/P_{\rm orb}$, where $25\,\mbox{d}$ is taken as the duration over which data is collected in a single sector. The factor of two comes from the fact that on average, each star has TESS-SPOC light curves from two sectors. Given this probability, it is determined whether or not the transit is observed. If it is not, the object is recorded as a non-SLB. 
    \item For objects with an observed transit, we generate light curve models for the eclipse and self-lensing components. Those with a net positive signal (amplitude $> 0$) are stored as SLBs, and others as non-SLBs. 
    \item This process is repeated with the original population until we have a total of $6\times10^6$ binaries, within which there are $\sim 5000-6000$ SLBs. 
    \item We run the light curves of the SLBs through our self-lensing detection pipeline, retaining only simulated systems detected with S/N $>$ 10 and best-fit with the inverted transit/self-lensing model. As in the first test, the central time of the pulse is chosen randomly during the observation.  
\end{enumerate}
In Figure \ref{fig:injected_signal_test_2}, we show the results of this test. We plot the number and fraction of objects in different groups binned by various stellar and orbital parameters. Starting from the leftmost column and moving to the right, the parameters are: apparent TESS magnitude $T$, absolute G band magnitude $M_{\rm G, 0}$, stellar mass $M_{\star}$, orbital period $P_{\rm orb}$, and RV amplitude $2K_{\star}$ (i.e. two times the semi-amplitude). 

In the top row, we plot the number of all simulated WD + MS binaries. The distributions of $T$, $M_{\rm G, 0}$, and $M_{\star}$ match the distributions of the sample of MS stars observed by TESS -- mostly faint stars with $T \gtrsim 7$ with the mass distribution from Figure \ref{fig:mass_dist}. As expected, the $P_{\rm orb}$ distribution is uniform between $100 - 1000\,$d, and accordingly, $2K_{\star}$ peaks towards lower values. 

The next row shows the number of SLBs. We see that for most parameters, the distributions generally reflect that of the entire WD + MS binary population. While the eclipse probability increases as $R_{\star}$ increases for a fixed $M_{\star}$, it decreases as $M_{\star}$ increases for a fixed $R_{\star}$ (since the semi-major axis gets larger). Given the mass-radius relation for MS stars, these factors roughly cancel each other out which at least in part explains why the mass distribution for SLBs is similar to that of the original population. In contrast, the number monotonically decreases with $P_{\rm orb}$ as the eclipse probability and duty cycle fall. 

In the third row, we plot the number of SLBs whose signals were recovered (i.e. detected and best fit with the self-lensing model), and below that, the fraction of those relative to the total number of SLBs in each bin. We find that on average, the recovered fraction is roughly constant with brightness, though there is more scatter between sectors below $T \sim 7$.  There is a bigger contribution from stars with higher $M_{\rm G, 0}$ and lower $M_{\star}$ for which the self-lensing signals are stronger (Section \ref{ssec:LC_models}). However, the detection probability flattens out at the lowest $M_{\star}$ which may be due to these stars having more variability in their light curves as a result of stellar activity and star spots. For $P_{\rm orb}$, SLBs with longer periods are more likely to be detected if a self-lensing pulse occurs during TESS observations. 

Lastly, we plot the fraction of the same stars but over the total number of all WD + MS binaries in each bin. This value gives us the conversion between the number of SLBs that we find in our search and all AU-scale WD + MS binaries (Section \ref{sec:rates}). We find that in total, this fraction is $\sim 0.025\%$ across all sectors, assuming two sectors of observations per source and averaging over all periods and companion star masses and brightnesses.

\begin{figure*}
    \centering
    \includegraphics[width=0.9\textwidth]{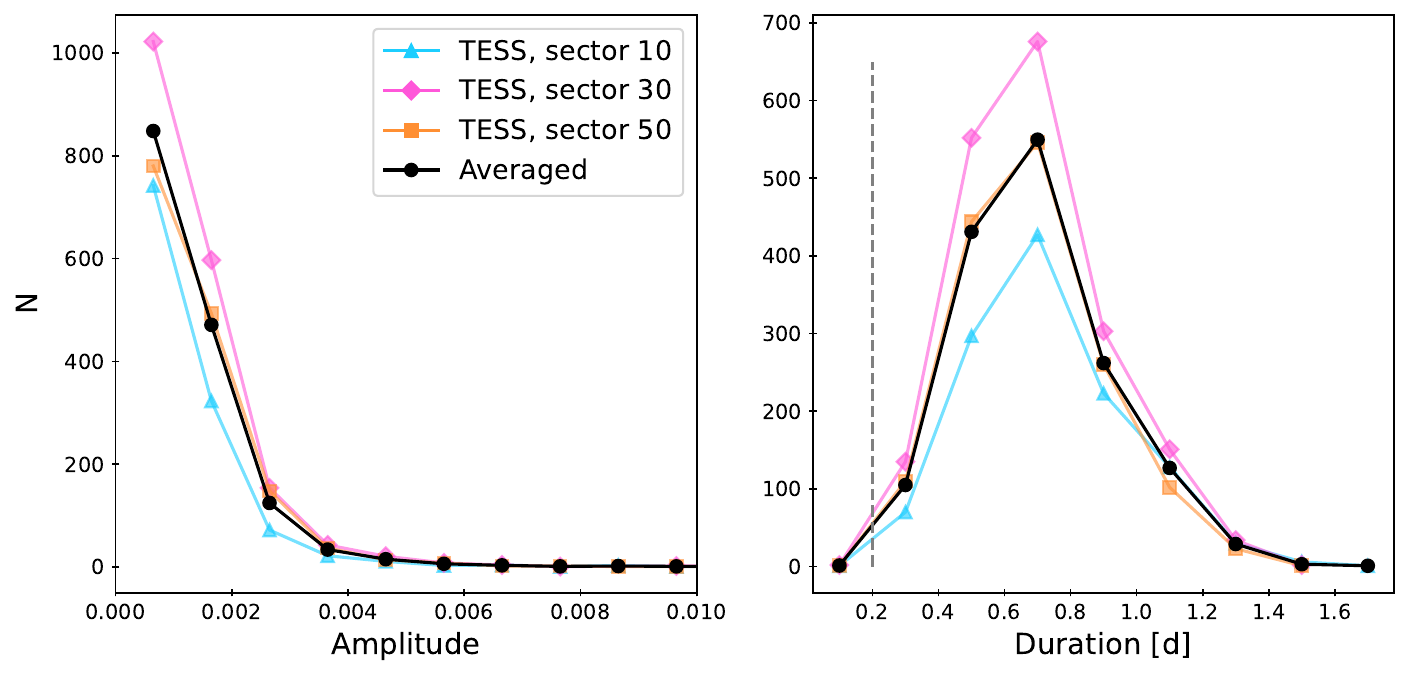}
    \caption{The number of injected signals recovered by our search pipeline, binned by the amplitude (\textit{Left}) and duration (\textit{Right}). The gray vertical line at a duration of $0.2\,$d is the lower limit that we set in narrowing down candidates. We also set an upper limit in the amplitude of 0.04 (Section \ref{sec:main_search}). We see that these cuts should not remove a significant fraction of self-lensing signals.}
    \label{fig:injected_signal_test_3}
\end{figure*}

In Figure \ref{fig:injected_signal_test_3}, we plot the number of signals best fit by an inverted transit model, binned by their amplitude and duration. We see that almost all self-lensing signals have amplitudes $\lesssim 0.005 $ and the majority have duration $> 0.2\,\mbox{d}$, meaning only a small fraction of SLBs would have been removed by the cuts we imposed on these quantities (Section \ref{sec:main_search}). 

We also carried out the same injection and recovery test for Kepler long-cadence light curves from quarter 5. A figure showing the results of this test, similar to Figure \ref{fig:injected_signal_test_2}, can be found in Figure \ref{fig:injected_signal_test_4} in Appendix \ref{appendix:kepler_inj_rec}. We find similar trends with respect to the different stellar and orbital parameters, but about an order of magnitude higher overall detection rate of $\sim 0.25\%$ compared to TESS. This is primarily due to the 4 year baseline of the observations -- since we are considering periods of up to 1000 days, all self-lensing binaries will have at least one observed pulse in this time. 
    
\section{Occurrence rates of AU-scale WD + MS binaries} \label{sec:rates}

Using the results of our injection and recovery test (Section \ref{ssec:injection_and_recovery_2}), we can estimate the fraction of all stars in AU-scale WD + MS binaries, $f_{\rm AU}$, based on the number of SLBs discovered from our search. The details of the following derivation are described in \citet{Green2024arXiv}, but we summarize the key steps here. 

The probability of detecting a self-lensing pulse from a given WD + MS binary depends on its stellar and orbital parameters: 
\begin{equation}
    p\left(\mbox{detect}|\textbf{\textit{Y}}, P_{\rm orb}, \mbox{cos}i \right) = S\left(\textbf{\textit{Y}}, P_{\rm orb}, \mbox{cos}i \right)
\end{equation}
where $\textbf{\textit{Y}}$ is a vector of parameters describing the MS star, which we assume to be independent of the orbit, and $S$ represents the selection function of our search, which we estimated via simulations above. We do not consider the dependence on $M_{\rm WD}$ as we fixed it to be $0.6\,M_{\odot}$ in our test. 

Integrating this over the binary parameters, we get the probability of detecting self-lensing for a binary hosting a given primary: 
\begin{equation}
    p(\mbox{detect}|\textbf{\textit{Y}}) = \int d\mbox{cos}i \int dP_{\rm orb} \ p(P_{\rm orb}) p(\mbox{cos}i) S
\end{equation}
where $p(P_{\rm orb})$, and $p(\mbox{cos}i)$ are the assumed prior distributions on the respective parameters, which in our case are uniform. 

The expected number of detected SLBs is: 
\begin{equation}
    N_{\rm SLB, found} = f_{\rm AU} \sum_{j = 1}^{N_{\rm input}} p(\mbox{detect}|\textbf{\textit{Y$_j$}}) = f_{\rm AU} N_{\rm input} \bar{\mathcal{S}}
\end{equation}
where $N_{\rm input}$ is the total number of MS stars in our input sample, $f_{\rm AU}$ is the fraction that we would like to calculate (modeled to be constant over $P_{\rm orb} = 100 - 1000\,$d), and $\bar{\mathcal{S}}$ is the SLB detection efficiency of our algorithm averaged over all stellar and binary parameters. $\bar{\mathcal{S}}$ can be approximated by the fraction of SLBs that make it into our sample over all simulated WD + MS binaries, which we found in Section \ref{sec:injection_and_recovery}. This approximation is valid because the simulated population was drawn from the desired distributions $p(\cos i)$ and $p(P_{\rm orb})$, and the distribution of $\textbf{\textit{Y}}$ used in the simulations matches the observed sample.

\subsection{Rates based on self-lensing}

Kepler observed $\sim 10^5$ stars and from these, 5 SLBs were identified. The eclipse probability is the ratio of the stellar radius to the size of the orbit, which for the objects in \citet{Kawahara2018AJ}, is approximately $1/200$. Thus, a simple geometric estimate is that $\sim 100 \cdot (5 \cdot  200)/10^5 = 1 \%$ of all stars have WD companions in AU-scale orbits \citep{Masuda2019ApJL, Masuda2020IAUS}. Here, we perform a more careful calculation using the results of our injection and recovery tests. 

We ran the Kepler light curves of the five known SLBs through our algorithm and found that at least one self-lensing signal for four of these were detected. KOI-3278 was missed due to the presence of star spots which lead to quasi-periodic variability in its light curves \citep{Kruse2014Sci}. Our pipeline does not effectively de-trend such variability, which resulted in poor fits of our self-lensing model. 

From the four SLBs which were recovered by our pipeline, we can calculate an estimate of the occurrence rate -- more precisely, this is a lower limit since we did not search all Kepler light curves and it is possible other SLBs were missed in previous searches. We found in Section \ref{ssec:injection_and_recovery_2} that $\bar{\mathcal{S}} \sim 0.25\%$ for Kepler. Given that it observed around 142,000 MS stars and found $4 \pm 2$ SLBs (where the uncertainty comes from the Poisson error), we find that $f_{\rm AU} \sim 100 \cdot 4/(142000 \cdot 0.0025) = (1.1 \pm 0.6) \%$ of all MS stars orbit WDs with $P_{\rm orb} \sim 100 - 1000\,$d. Note that here, we do not try to incorporate the uncertainty coming from our measurement of $\bar{\mathcal{S}}$ which depends on factors such as the assumed period and mass distributions and is beyond the scope of this work. 

Meanwhile for TESS, we found in Section \ref{ssec:injection_and_recovery_2} that $\bar{\mathcal{S}} \sim 0.025\%$. Since there are about 430,000 MS stars in our original sample, this tells us that if we find 1 SLB, $f_{\rm AU} \sim 100 \cdot 1/(430,000 \cdot 0.00025) \sim 1\%$. This is roughly consistent with the above estimate from Kepler, meaning that if little were missed in those searches, we expect to find $\mathcal{O}(1)$ SLBs in our search.

\subsection{Comparison to Gaia MS + WD binaries}

Recently, \citet{Shahaf2024MNRAS} identified over 3000 astrometric WD + MS binary candidates using data from Gaia DR3. These have similar orbital periods to the SLBs, ranging from $\sim 50 - 1000\,$d with a median at $\sim 650\,$d. These AU-scale MS + WD binaries are of interest in the study of binary mass transfer. They have orbital separations that are too close to have avoided mass transfer when the WD progenitor became a giant. Meanwhile, their separations are both too small and too wide to conclusively be post- stable mass transfer or post- common envelope systems \citep[e.g.][]{Belloni2024A&A_2, Yamaguchi2024PASP, Yamaguchi2024PASP_2}. Measurements of their intrinsic space densities and occurrence rates are important to make comparisons to theoretical predictions. 

However, due to the multi-step selection process implemented by Gaia in choosing objects that get astrometric binary solutions \citep{Halbwachs2023A&A}, the selection function for this sample is complex, making it challenging to obtain an intrinsic rate. Nevertheless, we can make an order of magnitude estimate to compare to the values calculated in the previous section. 

\begin{figure*}
    \centering
    \includegraphics[width=0.98\textwidth]{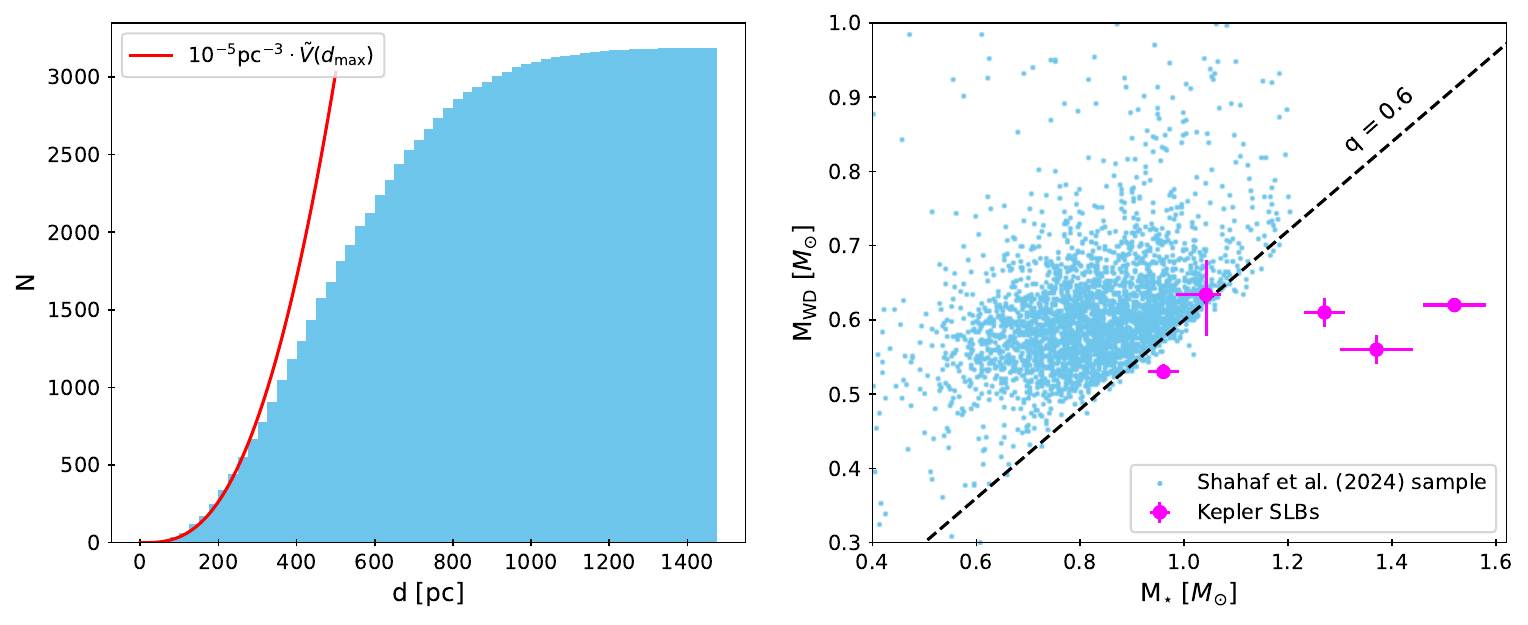}
    \caption{\textit{Left}: Cumulative distance distribution of the \citet{Shahaf2024MNRAS} sample of astrometric WD + MS binary candidates. The red curve is the effective volume, $\tilde{V}$, enclosed in a distance, $d_{\rm max}$, multiplied by the midplane number density of $10^{-5}\,\mbox{pc}^{-3}$. \textit{Right}: WD mass against the luminous star mass for the \citet{Shahaf2024MNRAS} sample and the five known Kepler SLBs \citep{Kawahara2018AJ, Masuda2019ApJL, Masuda2020IAUS, Yamaguchi2024PASP}. The dashed line marks $q = M_{\rm WD}/M_{\star} = 0.6$. We see that the vast majority of the astrometric sample lie above this line which is a selection effect of the AMRF method. Meanwhile, all SLBs lie close or firmly below the line and therefore are unlikely to have been detected with this method, even if they had astrometry.}
    \label{fig:shahaf24_distance}
\end{figure*}

The cumulative distance distribution of the \citet{Shahaf2024MNRAS} sample is shown in the left panel of Figure \ref{fig:shahaf24_distance}. The red curve is the effective volume enclosed within a distance $d_{\rm max}$ from the Sun, $\tilde{V}\left(d_{\rm max}\right)$, multiplied by the midplane number density which we fit here by eye to be $\sim10^{-5}\,\mbox{pc}^{-3}$. We assume a plane-parallel exponential distribution for the number density with a scale height of $300\,$pc, which is taken into account by the definition of $\tilde{V}$ \citep{El-Badry2021MNRAS}. The implied midplane number density is roughly constant up to a distance of $\sim 200\,\mbox{pc}$, implying that the completeness is roughly constant within that volume. Given a local stellar density of $0.1\,\mbox{pc}^{-3}$ \citep[e.g.][]{Golovin2023A&A}, we estimate that $f_{\rm AU} \sim (1\times10^{-5})/(0.1) \cdot 100 = 0.01\%$ which is 2 orders of magnitude smaller than the 1\% fraction implied by SLBs. 

However, $0.1\,\mbox{pc}^{-3}$ is the number density of all stars in the solar neighbourhood and is dominated by M dwarfs. On the other hand, the Gaia and TESS samples are dominated by solar-type stars. Therefore, the above comparison comes with the caveat that it is assuming M dwarfs and solar-type stars are equally likely to have wide WD companions, which may not be true. For a more conservative estimate, we select stars from Gaia DR3 within $100\,$pc and with extinction-corrected absolute G-band magnitudes between 3 and 6 (corresponding to stars with masses $\sim 0.8 - 1.4\,M_{\odot}$), and estimate their number density using their cumulative distance distribution as we did previously. We find a midplane number density of solar-type stars of $0.008\,\mbox{pc}^{-3}$. The number density of WD+MS binaries from the \citet{Shahaf2024MNRAS} sample restricted to $M_{G,0}=3-6$ is $\sim 4\times10^{-6}\,\mbox{pc}^{-3}$, which corresponds to a wide WD companion fraction of $f_{\rm AU} \sim 0.05\%$ for solar-type stars in particular. 

While these are rough estimates, the fact that the inferred $f_{\rm AU}$ value based on the Gaia sample from \citet{Shahaf2024MNRAS} is at least an order of magnitude smaller than the fraction implied by SLBs suggests that the Gaia sample is highly incomplete. 

Such incompleteness could arise for at least two reasons. First, some binaries simply do not receive astrometric solutions due to a variety of choices made in the Gaia astrometric binary processing \citep[e.g.][]{El-Badry2024}. Second, some do receive astrometric solutions, but these do not allow the companion to be unambigously recognized as a WD. In particular, \citet{Shahaf2024MNRAS} selected systems that are likely to host WD companions based on their having sufficiently large values of the astrometric mass ratio function (AMRF). Only for systems that have mass ratios $q = M_{\rm WD}/M_{\star} \gtrsim 0.6$ can a luminous companion be ruled out \citep[e.g.][]{Hallakoun2024}. At least four out of the five known SLBs would not be included in this sample, even if they had astrometric orbits in Gaia DR3, as they have $q < 0.6$ or equivalently, AMRF values that are too small to rule out luminous companions. This is illustrated in the right panel of Figure \ref{fig:shahaf24_distance}. 

\section{Conclusion} \label{sec:conclusion}
In this work, we presented a method to search for self-lensing binaries by identifying non-repeating signals using data from TESS. We summarize the key takeaways below: 
\begin{itemize}
    \item While our sample of TESS-SPOC light curves is made up of a greater number of stars than Kepler observed in total, due to TESS being a smaller telescope with each star having a shorter baseline, we are less likely to observe a self-lensing event (Figure \ref{fig:lc_days}). On the other hand, our stars are on average closer and brighter, making them easier to follow up (Figure \ref{fig:mass_dist}). In the future, we will search through all TESS light curves, as opposed to just those from TESS-SPOC, which is expected to significantly increase discovery potential.
    \item Based on the method from \citet{Foreman-Mackey2016AJ}, we identify candidate self-lensing pulses based on a S/N cut, then select those with the expected inverted transit shape using Gaussian Process fitting to several models (Section \ref{sec:main_search}). We also carry out a separate search isolating sources with orbital solutions in Gaia DR3, as these will require less RV follow-up to constrain their orbits (Section \ref{ssec:nss_search}). 
    \item We provide a list of candidates after manual inspection with large RV amplitudes $> 10\,\mbox{kms}^{-1}$ and RUWEs $> 1.0$, possibly indicative of binarity at the expected orbital periods (Table \ref{tab:candidates_list}). 
    \item We simulate a population of WD + MS binaries, identifying SLBs, and injecting artificial signals into their light curves. We then run these through our pipeline to measure detection efficiency. This allows us to estimate the occurrence rate of AU-scale WD + MS binaries. Given the known Kepler SLBs, this tell us that $(1.1 \pm 0.6) \%$ of all stars are found in these systems and that we expect to find $\mathcal{O}(1)$ SLBs from our search in TESS. (Section \ref{ssec:injection_and_recovery_2})
    \item A simple estimate based on the distance distribution of astrometric WD + MS binaries from Gaia \citep{Shahaf2024MNRAS} leads to much lower occurrence rates of $\sim0.01-0.05\%$, suggesting that this sample is quite incomplete. This is primarily due to the selection criteria of \citet{Shahaf2024MNRAS} which require systems to have $q > 0.6$ to rule out luminous secondaries, making it insensitive to many WD SLBs with smaller mass ratios.  
\end{itemize}
We are currently obtaining spectroscopic follow-up of our best candidates and will present those results in future work. 

\section{Acknowledgements}

We thank Sterl Phinney and Jim Fuller for providing helpful comments on our manuscript. NY and KE acknowledge support from NSF grant AST-2307232. NY acknowledges support from the Ezoe Memorial Recruit Foundation scholarship. This research benefited from discussions in the ZTF Theory Network, funded in part by the Gordon and Betty Moore Foundation through Grant GBMF5076. 

This paper includes data collected by the TESS mission. Funding for the TESS mission is provided by the NASA's Science Mission Directorate. This work has made use of data from the European Space Agency (ESA) mission
{\it Gaia} (\url{https://www.cosmos.esa.int/gaia}), processed by the {\it Gaia}
Data Processing and Analysis Consortium (DPAC,
\url{https://www.cosmos.esa.int/web/gaia/dpac/consortium}). Funding for the DPAC
has been provided by national institutions, in particular the institutions
participating in the {\it Gaia} Multilateral Agreement.

\appendix

\section{A false positive} \label{appendix:TIC_55276301}

In an earlier version of our manuscript, we described a promising target from our search, TIC 55276301. After taking a closer look at its FFIs, it has been identified to be a faint asteroid crossing event. Nevertheless, we leave a description of its properties here for reference.

\begin{figure*}
    \centering
    \includegraphics[width=0.95\textwidth]{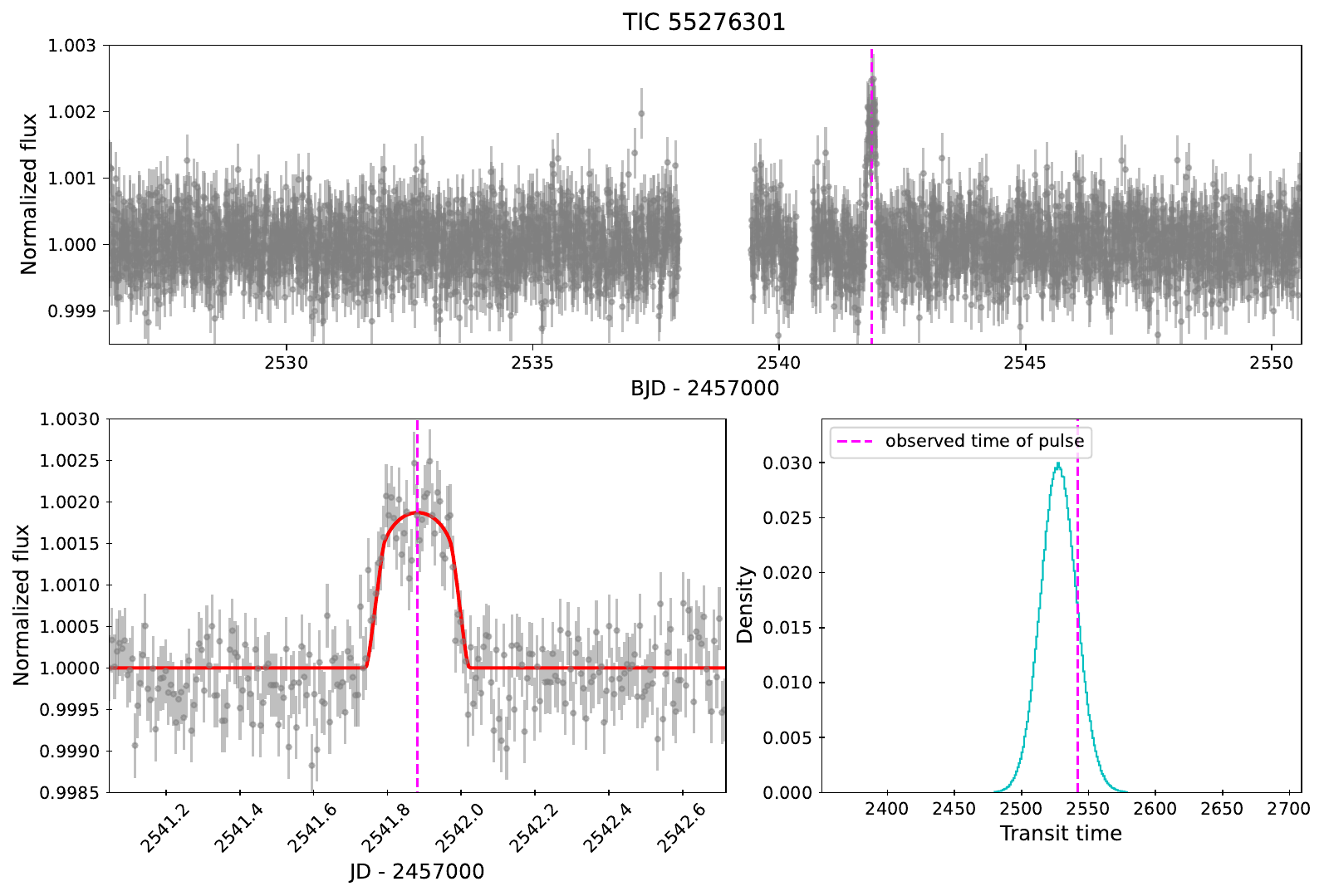}
    \caption{\textit{Top}: Candidate self-lensing signal from sector 45 of TIC 55276301. \textit{Bottom left}: Zoom-in onto the signal. In red, we plot our preliminary best fit self-lensing model with $M_{\star} = 1.08 M_{\odot}$ and $i = 89.73$. \textit{Bottom right}: Probability distribution of the expected transit time for this object based on the Gaia orbital solution. The range in the x-axis reflects the orbital period of this system (358 days). The magenta dashed lines mark the central time of the pulse, which is around $1\sigma$ away from the peak of the expected time.}
    \label{fig:TIC55276301}
\end{figure*}

The light curve of this object, and a zoom-in on the detected pulse, is shown on the top and bottom left panel of Figure \ref{fig:TIC55276301}. This object has an SB1 solution in Gaia DR3 (source ID 631849788720040448) with a $358 \pm 2$ day orbital period. It was identified in both the main and NSS-focused searches. 

By taking the Gaia orbital parameters, we can fit the pulse and get preliminary constraints on the luminous star mass, $M_{\star}$, and inclination, $i$, in the case it is a self-lensing binary. The radii of the stars are estimated using mass-radius relations \citep{Eker2018MNRAS, Nauenberg1972ApJ}, the limb-darkening coefficients are fixed ($u_1 = 0.4, u_2 = 0.3$), and the WD mass is from the mass function of the Gaia SB1 orbit along with the two fitted parameters. The resulting best-fit model is plotted in Figure \ref{fig:TIC55276301}, where $M_{\star} = 1.08 M_{\odot}$ and $i = 89.73^{\circ}$ for which $M_{\rm WD} = 0.69 M_{\odot}$. This value of $M_{\star}$ is reasonable given the absolute magnitude of the star ($M_G=3.8$; Section \ref{ssec:tess_properties}) and $M_{\rm WD}$ is within the expected range for a WD. 

Due to its almost circular orbit ($e \sim 0.02$), the periastron time is very poorly determined. On the bottom right panel of Figure \ref{fig:TIC55276301}, we show the dispersion in the expected transit time (i.e. time of conjunction) calculated from the Gaia solution. We see that the central time of the observed pulse is consistent with the Gaia constraints, but the uncertainty is large because the Gaia solution is based on data obtained 6-9 orbits prior to the TESS observation. 

While we have ruled out this source to be a SLB, we note that follow-up spectra show no contribution from a second luminous companion which would be expected for a MS star of mass $\gtrsim 0.7\,M_{\odot}$. Therefore, it is likely a WD + MS binary, though not one that is self-lensing.

\section{List of SLB candidates} \label{appendix:candidate_tables}

\begin{figure*}
    \centering
    \includegraphics[width=0.95\textwidth]{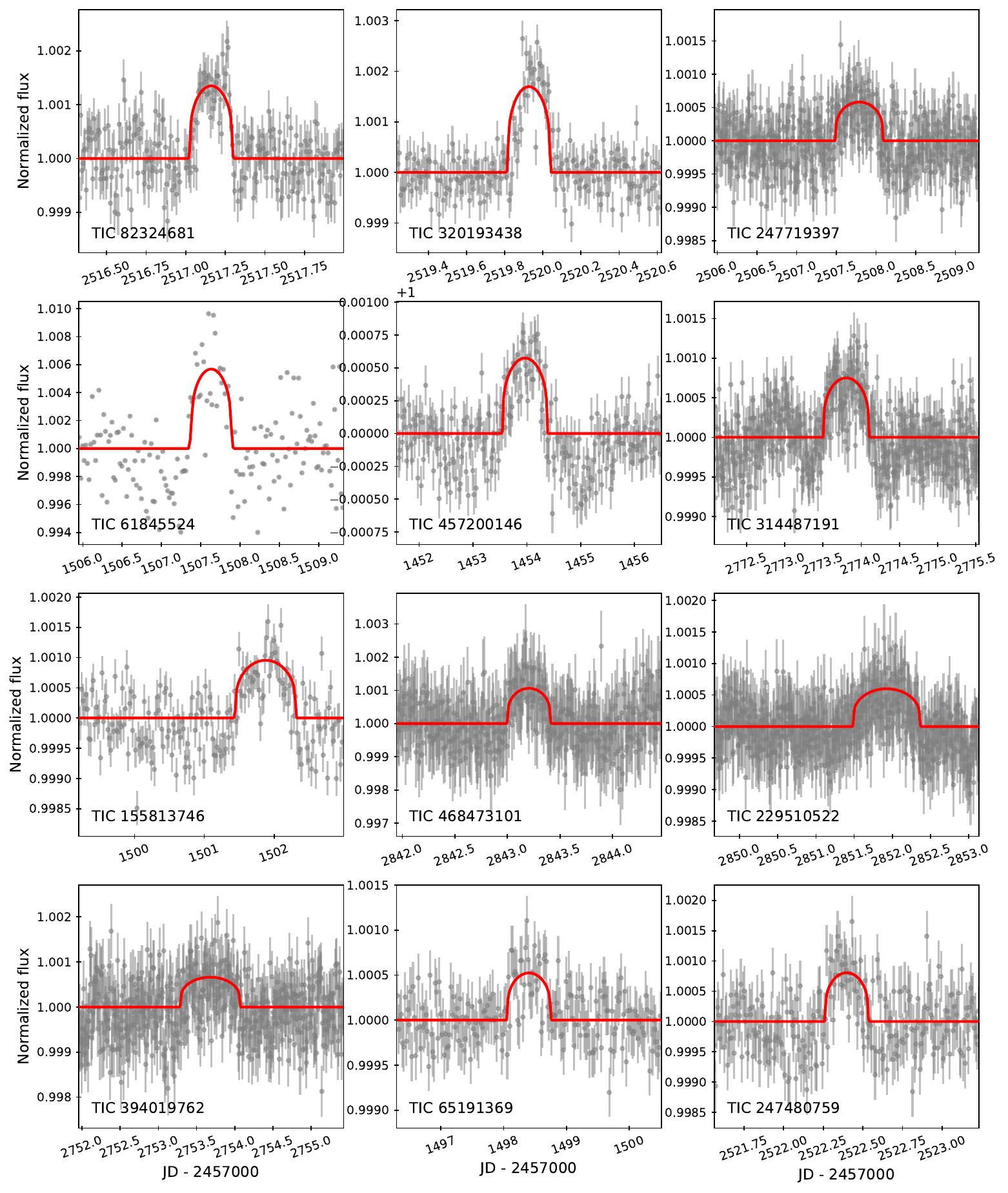}
    \caption{Objects from the top 12 rows of Table \ref{tab:candidates_list}. We plot the signals detected, along with the best-fit inverted transit model.}
    \label{fig:best_candidates}
\end{figure*}

In Table \ref{tab:candidates_list}, we list candidate SLBs from our main and focused NSS searches, and in Figure \ref{fig:best_candidates}, we show the light curves of a subset of these around the detected peaks (Section \ref{sec:main_search}).

\section{Kepler Injection and Recovery} \label{appendix:kepler_inj_rec}

\begin{figure*}
    \centering
    \includegraphics[width=0.99\textwidth]{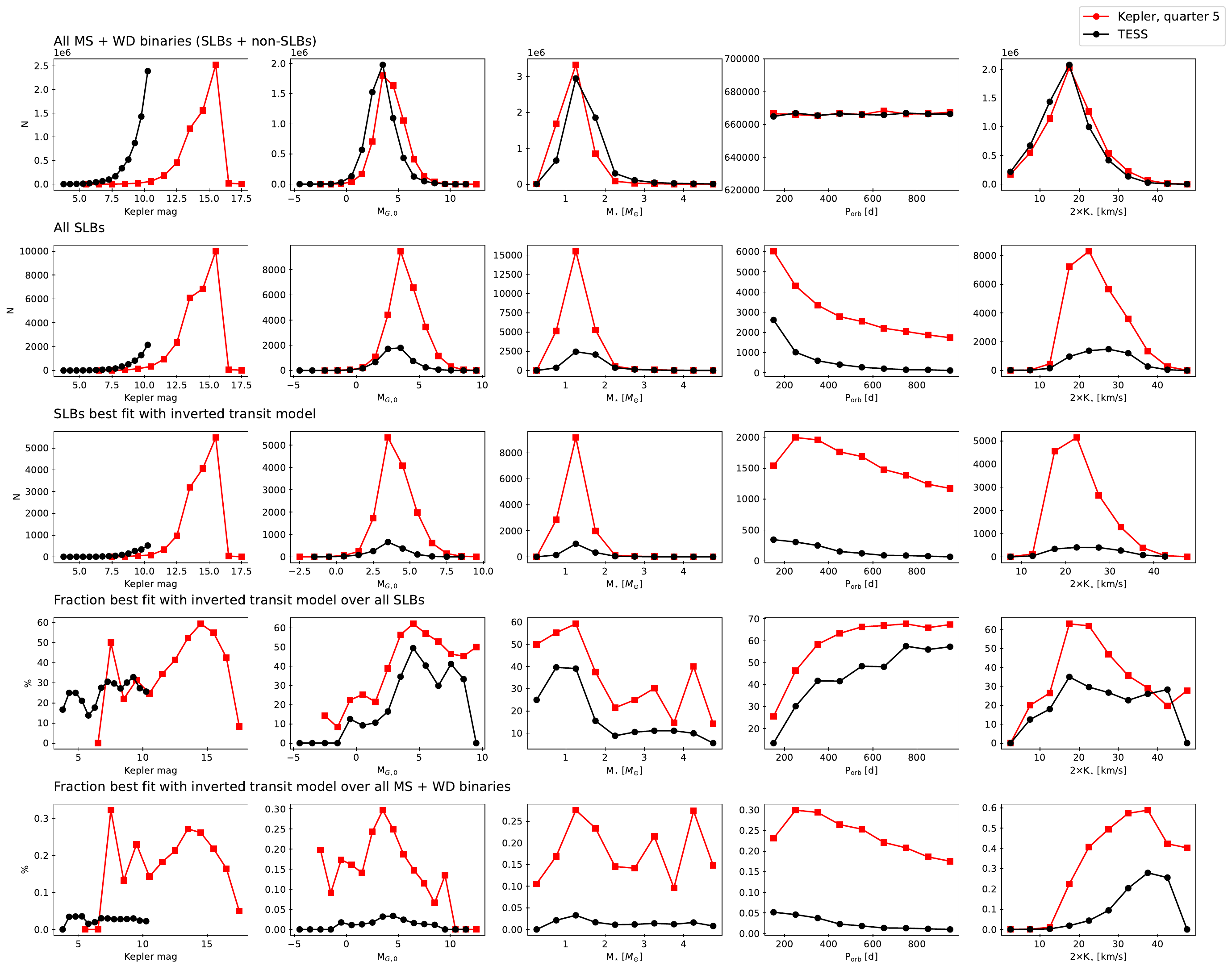}
    \caption{Results of injection and recovery tests using light curves from quarter 5 of Kepler (red). The black lines are the averaged curves for TESS from Figure \ref{fig:injected_signal_test_2}, where a description for each panel can be found.}
    \label{fig:injected_signal_test_4}
\end{figure*}

In Figure \ref{fig:injected_signal_test_4}, we show the results of performing injection and recovery tests on Kepler, compared to TESS, light curves (Section \ref{ssec:injection_and_recovery_2}). A detailed description of each panel can be found in the caption of Figure \ref{fig:injected_signal_test_2}.

\begin{longrotatetable}
\begin{deluxetable}{c c c c c c c c c c c c c}
\setlength\extrarowheight{7pt}
\tablecaption{Table of objects that made it into our final list of candidate SLBs, with RUWE $> 1.0$ and RV amplitude $> 10\,\mbox{km/s}$, as described in Section \ref{sec:main_search}. We also include several candidates from our focused search in the NSS catalog (Section \ref{ssec:nss_search}). We note their key characteristics from manual inspection, which aided in determining whether or not they were selected for spectroscopic follow-up. Note that ASB1 = AstroSpectroSB1, and CVZ = Continuous Viewing Zone.}
\tabletypesize{\scriptsize}
\tablehead{
\colhead{TIC} & \colhead{source ID} & \colhead{Search} & \colhead{Note} & \colhead{RUWE} & 
\colhead{RV amplitude} & \colhead{$M_G$} & 
\colhead{$M_{\star}$ [$M_{\odot}$]} & \colhead{NSS} & 
\colhead{$P_{\rm orb}$ [$M_{\odot}$]} & \colhead{$e$} & 
\colhead{$M_{\rm WD}$ [$M_{\odot}$]}
} 
\startdata
         82324681 & 657510191090704640 & main & \makecell[c]{Good fit, large RUWE and RV amp, \\ dip in Sector 46 caused by asteroid} & 6.04 & 24.77 & 4.12 & 1.19 & nan & nan & nan & nan \\
         320193438 & 669719821039909888 & main & \makecell[c]{Good fit, large RUWE and RV amp, \\ short duration}  & 4.39 & 23.60 & 3.87 & 1.26 & nan & nan & nan & nan \\
         247719397 & 3346301795905002368 & main & \makecell[c]{Pulse a little weak, somewhat crowded,\\ highly eccentric} & 1.04 & 17.62 & 2.38 & 1.65 & nan & nan & nan & nan \\
         61845524 & 3109952734908875520 & main & \makecell[c]{Pulse a little weak, somewhat crowded}  & 2.30 & 35.12 & 2.32 & 1.67 & SB1 & 147 & 0.43 & 0.58 \\
         457200146 & 3232916033679156224 & main & \makecell[c]{Fit okay, large RV amp,\\ but HB-like shape} & 1.13 & 23.49 & 1.33 & 2.32 & nan & nan & nan & nan \\
         314487191 & 4532321418003888128 & main & \makecell[c]{Fit okay, other variabiliy present} & 1.43 & 14.47 & 0.58 & 3.11 & nan & nan & nan & nan \\
         155813746 & 5692765712460857472 & main & \makecell[c]{Other variability present, \\large RV amp, very crowded} & 1.06 & 29.87 & 1.42 & 2.24 & nan & nan & nan & nan \\
         468473101 & 1965726168256488704 & main & \makecell[c]{Pulse a little weak, very crowded}  & 4.63 & 21.55 & 3.40 & 1.39 & nan & nan & nan & nan \\
         229510522 & 2247655071463769600 & main & \makecell[c]{Pulse a little weak, \\ CVZ but only one pulse} & 2.02 & 17.48 & 2.08 & 1.78 & nan & nan & nan & nan \\
         394019762 & 4543783242611185280 & main & \makecell[c]{Pulse a little weak, crowded} & 3.87 & 37.11 & 1.39 & 2.27 & nan & nan & nan & nan \\
         65191369 & 5611129516718902272 & main & \makecell[c]{Pulse a little weak, very crowded} & 2.04 & 47.09 & 3.35 & 1.41 & nan & nan & nan & nan \\
         247480759 & 3347433674406262656 & main & \makecell[c]{Fit okay, noisy, crowded, \\signal appears off source} & 1.41 & 17.70 & 5.39 & 0.92 & nan & nan & nan & nan \\
         230114166 & 1648830080850407552 & main & \makecell[c]{Fit okay, very large RV amp} & 1.12 & 88.01 & 2.31 & 1.67 & nan & nan & nan & nan \\
         124508784 & 3100778405583395072 & main & \makecell[c]{Fit not great, variability, crowded} & 1.48 & 20.23 & 1.91 & 1.87 & nan & nan & nan & nan \\
         299891691 & 6688849614297149824 & main & \makecell[c]{Fit not great, step-like} & 1.49 & 10.29 & 2.92 & 1.52 & nan & nan & nan & nan \\
         9293371 & 3181637010382678400 & main & \makecell[c]{Fit not great} & 1.00 & 20.93 & 1.54 & 2.15 & nan & nan & nan & nan \\
         369000430 & 4037702779416807552 & main & \makecell[c]{Two peaks but slightly \\different relative flux} & 1.54 & 11.65 & 0.75 & 2.91 & nan & nan & nan & nan \\
         29850601 & 5390373282936805376 & main & \makecell[c]{HB-like shape} & 1.15 & 34.08 & 1.85 & 1.91 & nan & nan & nan & nan \\
         307364849 & 5272990074390867840 & main & \makecell[c]{HB-like shape} & 1.12 & 14.60 & 1.56 & 2.13 & nan & nan & nan & nan \\
         234632282 & 3124127982407698560 & main & \makecell[c]{HB-like shape} & 1.09 & 27.43 & 1.30 & 2.35 & nan & nan & nan & nan \\
         151937435 & 5397005399634166784 & main & \makecell[c]{HB-like shape, crowded field} & 4.22 & 25.52 & 3.14 & 1.47 & nan & nan & nan & nan \\
         453368802 & 976007747235767168 & main & \makecell[c]{near edge of data gap} & 1.59 & 12.52 & 1.78 & 1.96 & nan & nan & nan & nan \\
         293164329 & 4775452026208341248 & main & \makecell[c]{near edge of data gap} & 10.64 & 15.36 & 5.58 & 0.89 & ASB1 & 584 & 0.28 & 0.34 \\
         238004509 & 5498660609742310272 & main & \makecell[c]{near edge of data gap} & 1.00 & 16.88 & 0.42 & 3.32 & nan & nan & nan & nan \\
         320193939 & 669027128714437120 & main & \makecell[c]{short pulse duration, period too short} & 1.17 & 42.39 & 4.16 & 1.18 & SB1 & 15 & 0.02 & 0.31 \\
         63119966 & 2125813235891371392 & nss & \makecell[c]{good fit, low WD mass,\\possibly background eclipse} & 4.47 & 11.63 & 1.75 & 1.98 & SB1 & 803 & 0.09 & 0.34 \\
\enddata
\end{deluxetable} \label{tab:candidates_list}
\end{longrotatetable}

\bibliographystyle{aasjournal}

\end{document}